\documentclass{MyPoS}

\usepackage{here}
\usepackage{graphicx}
\usepackage{epsfig}

\newcommand{\no}{\noindent}
\newcommand{\myeq}[3]{\vspace{#2} \begin{equation} \hspace{#1} #3 \end{equation} \vspace{0cm}}
\newcommand{\myeqn}[3]{\vspace{#2} \begin{displaymath} \hspace{#1} #3 \end{displaymath} \vspace{0cm}}
\newcommand{\vsp}[1]{\vspace*{#1}}
\newcommand{\hsp}[1]{\hspace*{#1}}
\newcommand{\Op}{\mathcal{O}}

\title{Constraints on low energy QCD parameters from $\eta \to 3\pi$ and $\pi\pi$ scattering}

\ShortTitle{Extraction of low energy QCD parameters from $\eta \to 3\pi$ and beyond}

\author{Mari\'{a}n Koles\'{a}r\\
        Institute of Particle and Nuclear Physics, Faculty of Mathematics and Physics, Charles University in Prague, 
				CZ-18000 Prague, Czech republic\\
        E-mail: \email{kolesar@ipnp.troja.mff.cuni.cz}}

\author{Ji\v{r}\'i Novotn\'y\\
				Institute of Particle and Nuclear Physics, Faculty of Mathematics and Physics, Charles University in Prague, 
				CZ-18000 Prague, Czech republic\\
        E-mail: \email{novotny@ipnp.troja.mff.cuni.cz}}

\abstract{The $\eta$$\,\to\,$$3\pi$ decays are a valuable source of information on low energy QCD. Yet they were not used for an extraction of the three flavor chiral symmetry breaking order parameters until now. We use a Bayesian approach in the framework of resummed chiral perturbation theory to obtain constraints on the quark condensate and pseudoscalar decay constant in the chiral limit. We compare our results with recent CHPT and lattice QCD fits and find some tension, as the $\eta$$\,\to\,$$3\pi$ data seem to prefer a larger ratio of the chiral order parameters. The results also disfavor a very large value of the pseudoscalar decay constant in the chiral limit, which was found by some recent works. In addition, we present results of a combined analysis including $\eta$$\,\to\,$$3\pi$ decays and $\pi\pi$ scattering and though the picture does not changed appreciably, we find some tension between the data we use. We also try to extract information on the mass difference of the light quarks, but the uncertainties prove to be large.}

\begin{document}

\section{Introduction}

Spontaneous breaking of chiral symmetry (SB$\chi$S) is a prominent feature of the QCD vacuum and thus its character has been under discussion for a long time \cite{Fuchs:1991cq,DescotesGenon:1999uh}. The principal order parameters are the quark condensate and the pseudoscalar decay constant in the chiral limit\footnote{We will abbreviate these to \emph{chiral condensate} and \emph{chiral decay constant} in the following}

\myeq{0cm}{0cm}{\label{Sigma}
	 \Sigma(N_f) = -\langle\,0\,|\,\bar{q}q \,|\,0\,\rangle\,|_{m_q\to 0}\,,}
\myeq{0cm}{-0.25cm}{\label{F}
	 F(N_f) = F_P^a\,|_{m_q\to 0}\,,\quad   i p_{\mu}\, F_P^a\ =\ \langle\,0\,|\,A_{\mu}^a\,|\,P\,\rangle,}

\no where $N_f$ is the number of quark flavors $q$ considered light and $m_q$ collectively denotes their masses. $A_{\mu}^a$ are the QCD axial vector currents, while $F_P^a$ the decay constants of the light pseudoscalar mesons $P$. The two flavor parameters are usually denoted as $\Sigma$
and $F$, while the three flavor ones as $\Sigma_0$ and $F_0$.

Chiral perturbation theory ($\chi$PT) \cite{Weinberg:1978kz,Gasser:1983yg,Gasser:1984gg} is constructed as a general low energy parametrization of QCD based on its symmetries and the discussed order parameters appear at the lowest order of the chiral expansion as low energy constants (LECs). Interactions of the light pseudoscalar meson octet, the pseudo-Goldstone bosons of the broken symmetry, directly depend on the pattern of SB$\chi$S and thus can provide information about the values of $\Sigma(N_f)$ and $F(N_f)$.

A convenient reparametrization of these order parameters, relating them to physical quantities connected with pion two point Green functions, can be introduced \cite{DescotesGenon:1999uh}

\myeq{0cm}{0cm}{
	Z(N_f) = \frac{F(N_f)^2}{F_{\pi}^2},\quad
	X(N_f) = \frac{2\hat{m}\,\Sigma(N_f)}{F_{\pi}^2M_{\pi}^2},}

\no where $\hat{m}\,$=$\,(m_u\,$+$\,m_d)/2$. Defined in this way, $X(N_f)$ and $Z(N_f)$ are limited to the range (0,\,1). $Z(N_f)$\,=\,0 would correspond to a restoration of chiral symmetry and $X(N_f)$\,=\,0 to a case with vanishing chiral condensate. Standard approach to chiral perturbation series tacitly assumes values of $X(N_f)$ and $Z(N_f)$ not much smaller than one, which means that the leading order terms should dominate the expansion.

Several recent results for the two and three flavor order parameters are listed in Tables \ref{tab1} and \ref{tab2}, respectively. As can be seen, while the two flavor case is quite settled, the values of $X(2)$ and $Z(2)$ indeed being not much smaller than one, the situation in the three flavor sector is much less clear. Some analyses suggest a significant suppression of X(3) and/or Z(3) and thus a non-standard behavior of the spontaneously broken QCD vacuum. It can also be noted that the lattice averaging group FLAG \cite{Aoki:2016frl} does not report an average for the three flavor chiral order parameters. 

\begin{table}[b] \small \begin{center}
\begin{tabular}{|c|c|c|c|}
	\hline \rule[-0.2cm]{0cm}{0.5cm} phenomenology & $Z(2)$ & $X(2)$ \\
	\hline \rule[-0.2cm]{0cm}{0.5cm} $\pi\pi$ scattering \cite{DescotesGenon:2001tn} & 0.89$\pm$0.03 & 0.81$\pm$0.07 \\
	\hline \hline \rule[-0.2cm]{0cm}{0.5cm} lattice QCD & $Z(2)$ & $X(2)$ \\
	\hline \rule[-0.2cm]{0cm}{0.5cm} RBC/UKQCD+Re$\chi$PT \cite{Bernard:2012fw} & 0.86$\pm$0.01 & 0.89$\pm$0.01 \\
	\rule[-0.2cm]{0cm}{0.5cm} FLAG\,2016 $N_f$=2 \cite{Aoki:2016frl} & 0.87$\pm$0.02 & 0.84$\pm$0.11 \\
	\rule[-0.2cm]{0cm}{0.5cm} FLAG\,2016 $N_f$=2+1 \cite{Aoki:2016frl} & 0.88$\pm$0.01 & 0.86$\pm$0.03 \\
	\hline
\end{tabular} \end{center}
	\caption{Chosen results for the two flavor order parameters.}
	\label{tab1}
\end{table}\normalsize

\begin{table} \small \begin{center}
\begin{tabular}{|c|c|c|c|}
	\hline \rule[-0.2cm]{0cm}{0.5cm} phenomenology & $Z(3)$ & $X(3)$ \\
	\hline \rule[-0.2cm]{0cm}{0.5cm} NNLO $\chi$PT (BE14) \cite{Bijnens:2014lea} & 0.59 & 0.63\\
	\rule[-0.2cm]{0cm}{0.5cm} NNLO $\chi$PT (free fit) \cite{Bijnens:2014lea} & 0.48 & 0.45 \\
	\rule[-0.2cm]{0cm}{0.5cm} NNLO $\chi$PT ("fit 10") \cite{Amoros:2001cp} & 0.89 & 0.66 \\
	\rule[-0.2cm]{0cm}{0.5cm} Re$\chi$PT $\pi\pi$+$\pi K$ \cite{DescotesGenon:2007ta} & $>$0.2 & $<$0.8 \\
	\hline \hline \rule[-0.2cm]{0cm}{0.5cm} lattice QCD & $Z(3)$ & $X(3)$ \\
	\hline \rule[-0.2cm]{0cm}{0.5cm} RBC/UKQCD+Re$\chi$PT \cite{Bernard:2012ci} & 0.54$\pm$0.06 & 0.38$\pm$0.05\\
	\rule[-0.2cm]{0cm}{0.5cm} RBC/UKQCD+large $N_c$ \cite{Ecker:2013pba} & 0.91$\pm$0.08 & \\
	\rule[-0.2cm]{0cm}{0.5cm} MILC 09A \cite{Bazavov:2009fk} & 0.72$\pm$0.06 & 0.62$\pm$0.07 \\
	\hline
\end{tabular} \end{center}
	\caption{Chosen results for the three flavor order parameters.}
	\label{tab2}
\end{table}\normalsize

Up, down and strange quark masses are other parameters with strong influence on low energy QCD physics. A commonly used reparametrization can be introduced

\myeq{0cm}{0cm}{
	\hat{m} = \frac{m_u + m_d}{2},\quad
	r = \frac{m_s}{\hat{m}},\quad
	R = \frac{m_s-\hat{m}}{m_d-m_u}.}

\no The values for the light quark mass average and the strange to light quark mass ratio are well known from lattice QCD and QCD sum rules \cite{Aoki:2016frl,Narison:2014vka}. On the other hand, as can be seen in Table \ref{tab3}, the isospin breaking parameter $R$, directly related to the light quark mass difference, has not been determined with such a high precision by these or other methods yet.

\begin{table}[h] \small \begin{center}
\begin{tabular}{|c|c|c|}
	\hline \rule[-0.2cm]{0cm}{0.5cm} phenomenology & $R$ \\
	\hline \rule[-0.2cm]{0cm}{0.5cm} Dashen's theorem LO \cite{Bijnens:2007pr} & 44 \\
	\rule[-0.2cm]{0cm}{0.5cm} Dashen's theorem NNLO \cite{Bijnens:2007pr} & 37 \\
	\rule[-0.2cm]{0cm}{0.5cm} $\eta\to3\pi$ NNLO $\chi$PT \cite{Bijnens:2007pr} & 41.3  \\
	\rule[-0.2cm]{0cm}{0.5cm} $\eta\to3\pi$ dispersive \cite{Kampf:2011wr} & 37.7$\pm$2.2 \\
	\rule[-0.2cm]{0cm}{0.5cm} $\eta\to3\pi$ dispersive \cite{Colangelo:2016jmc} & 34.2$\pm$2.2 \\
	\rule[-0.2cm]{0cm}{0.5cm} $\eta\to3\pi$ dispersive \cite{Albaladejo:2017hhj} & 32.7$\pm$3.0 \\
	\hline \hline \rule[-0.2cm]{0cm}{0.5cm} lattice QCD & $R$ \\
	\hline \rule[-0.2cm]{0cm}{0.5cm}  FLAG\,2016 $N_f$=2 \cite{Aoki:2016frl} & 40.7$\pm$4.3 \\
	\rule[-0.2cm]{0cm}{0.5cm} FLAG\,2016 $N_f$=2+1 \cite{Aoki:2016frl} & 35.7$\pm$2.6 \\
	\hline
\end{tabular} \end{center}
	\caption{Chosen results for the isospin breaking parameter $R$. 
		$r=37.3\pm0.34$ \cite{Aoki:2016frl} was used to obtain $R$ from $Q$ 
		in ref.\cite{Colangelo:2016jmc} and \cite{Albaladejo:2017hhj}.}
	\label{tab3}
\end{table}\normalsize

In this paper, we use a Bayesian approach in the framework of resummed chiral perturbation theory to extract information on the three flavor chiral condensate, chiral decay constant and the mass difference of the light quarks. Our experimental input are well known observables connected to $\eta$$\,\to\,$$3\pi$ decays and $\pi\pi$ scattering. We assume a reasonable convergence of Green functions connected to these observables and investigate the constraints this assumption can provide for the discussed parameters. The results presented here are a significant update on our initial reports \cite{Kolesar:2013ywa,Kolesar:2014zra,Kolesar:2016iyz}

In Section \ref{sect_resummed_chpt} we shortly summarize our theoretical foundation. Section \ref{eta3pi_decays} discusses the $\eta$$\,\to\,$$3\pi$ decays, while Section \ref{Calculation} provides an overview of our calculation of these processes. Section \ref{sect_pipi_scattering} introduces $\pi\pi$ scattering into our analysis. The Bayesian statistical approach is reviewed in Section \ref{bayesian_analysis} and a discussion of our assumptions can be found in Section \ref{assumptions}. Section \ref{subthreshold_parameters} is concerned with investigating the compatibility of our theoretical predictions with the $\pi\pi$ scattering data. We employ a $\chi^2$ based analysis in Section \ref{chi2_analysis} to evaluate the quality with which our theoretical predictions reconstruct the experimental data and use it to choose between several
assumptions. Finally, in Section \ref{Results}, the main results of the Bayesian analysis are presented and compared with available literature. We conclude in Section \ref{Conclusions}.

\section{Resummed $\chi$PT \label{sect_resummed_chpt}}

We use an alternative approach to chiral perturbation theory, dubbed resummed $\chi$PT (Re$\chi$PT) \cite{DescotesGenon:2003cg}, which was developed in order to accommodate the possibility of irregular convergence of the chiral expansion. This is a typical scenario if the $X(3)$ and $Z(3)$ were indeed suppressed and the leading order was not dominant in the chiral expansion. In such a case one would have to be careful in the way how chiral expansion is defined and dealt with, as reshuffling of chiral orders could lead to unexpectedly large higher orders.     

The procedure can be shortly summarized in the following way:

\begin{itemize}	
	\item	The Standard $\chi$PT Lagrangian and power counting 
				\cite{Weinberg:1978kz,Gasser:1983yg,Gasser:1984gg} is used. In particular, the quark 
				masses $m_{q}$ are counted as $O(p^{2})$. \vsp{-0.25cm}
	\item	Only expansions of quantities related linearly to Green functions of QCD currents are 
				trusted, these are called \emph{safe observables}. It is assumed that the 
				next-to-next-to-leading and higher orders in these expansions are reasonably small, 
				though not necessary negligible. Leading order terms do not need to be dominant. 
				\vsp{-0.25cm}
	\item Calculations are performed explicitly to next-to-leading order, higher orders are 
				included implicitly in \emph{remainders}. The first step consists of performing the 
				\emph{strict chiral expansion} of the safe observables, which is understood as an expansion
				constructed in terms of the parameters of the chiral Lagrangian, while strictly respecting
				the chiral orders.\vsp{-0.25cm}
	\item In the next step, the strict expansion is modified in order to correct the location of 
				the branching points of the non-analytical part of the amplitudes, which need 
				to be placed in their physical position. This can be done either by means of a matching 
				with a dispersive representation or by hand. The \emph{bare expansion} is obtained.
				\vsp{-0.25cm}
	\item After that, an algebraically exact non-perturbative reparametrization of the 
				bare expansion is performed. It is obtained by expressing the $O(p^{4})$ LECs $L_i$ in 
				terms of physical values of experimentally well established safe 
				observables - the pseudoscalar decay constants and masses. The procedure generates 
				additional higher order remainders. We refer to these as \emph{indirect}
				remainders.\vsp{-0.25cm}
	\item The physical amplitude and other relevant observables are then obtained as algebraically 
				exact non-perturbative expressions in terms of the related safe observables and higher 
				order remainders.\vsp{-0.25cm}
	\item	The higher order remainders are not neglected, but estimated and treated as sources of 	
				error.
\end{itemize}

\no The hope for resummed $\chi$PT is that by carefully avoiding dangerous manipulations a better converging series can be obtained. The procedure also avoids the hard to control NNLO LECs by trading them for remainders with known chiral order.

In general, a chiral expansion of a safe observable is written in the following way:

\myeq{0cm}{0cm}{G = G^{(2)} + G^{(4)} + \Delta_G^{(6)} = G^{(2)} + G^{(4)} + G\,\delta_G^{(6)},}

\no where $\Delta_G^{(6)}= G\,\delta_G^{(6)}$ is the remainder which contains all terms starting with NNLO. The basic assumption of this paper is that $\delta_G^{(6)}\ll 1$ for our chosen observables. We will quantify this assumption later on.

\section{$\eta\to 3\pi$ decays \label{eta3pi_decays}}

The $\eta$$\,\to\,$$3\pi$ isospin breaking decays have not been exploited for an extraction of the chiral order parameters so far, yet we argue there is valuable information to be had. The theory seems to converge slowly for the decays. One loop corrections were found to be very sizable \cite{Gasser:1984pr}, the result for the decay width of the charged channel was 160$\pm$50 eV, compared to the current algebra prediction of 66 eV. However, the experimental value is still much larger, the current PDG value is \cite{PDG:2016xmw}

\myeq{0cm}{0cm}{\Gamma^+_\mathrm{exp} = 300 \pm 12 \ \mathrm{eV}.}

\no The latest experimental value for the neutral decay width is \cite{PDG:2016xmw}

\myeq{0cm}{0cm}{\Gamma^0_\mathrm{exp} = 428 \pm 17 \ \mathrm{eV}.}

\no Only the two loop $\chi$PT calculation \cite{Bijnens:2007pr} has succeeded to obtain a reasonable result for the widths.

As we have shown in \cite{Kolesar:2016jwe}, we argue that the four point Green functions relevant for the $\eta$$\,\to\,$$3\pi$ amplitude (see (\ref{Green_f}) below) do not necessarily have large contributions beyond next-to-leading order and a reasonably small higher order remainder is not in contradiction with huge corrections to the decay widths. The widths do not seem to be sensitive to the details of the Dalitz plot distribution, but rather to the value of leading order parameters - the chiral decay constant, the chiral condensate and the difference of $u$ and $d$ quark masses, i.e. the magnitude of explicit isospin breaking. Moreover, access to the values of these quantities is not screened by EM effects, as it was shown that the electromagnetic corrections up to NLO are very small \cite{Baur:1995gc, Ditsche:2008cq}. This is our motivation for our effort to extract information about the character of the QCD vacuum from this decay.

The Dalitz plot distributions are experimentally well known as well \cite{KLOE:2008ht,KLOE:2010mj,WASA:2014aks,BESIII:2015cmz,KLOE:2016qvh}. The usual parametrization of the square of the amplitude is defined as 

\myeq{0cm}{0cm}{
	|A(s,t;u)|^{2}=|A(s_{0},s_{0};s_{0})|^{2}
	\left(1+ay+by^{2}+dx^{2}+fy^{3}+gx^{2}y+\ldots \right)}

\no in the charged decay channel and as

\myeq{0cm}{0cm}{
		|\overline{A}(s,t;u)|^{2}=|A(s_{0},s_{0};s_{0})|^{2}
		\left( 1+2\alpha z+\ldots \right)}
	
\no in the $\eta\to 3\pi^0$ case, where	
		
\myeq{0cm}{0cm}{
	x = -\frac{\sqrt{3}\left( t-u\right)}{2M_{\eta}(M_\eta-3M_\pi)}, \quad
	y = -\frac{3\left(s-s_{0}\right)}{2M_{\eta }(M_\eta-3M_\pi)} \quad
	z = x^{2}+y^{2}}

\no and $s_0=\frac{1}{3}(s+t+u)$ is the center of the Dalitz plot.

However, as we have discussed in detail in \cite{Kolesar:2016jwe}, we have not found the convergence of the theory in the case of the slopes reliable enough to include all the experimentally measured Dalitz plot parameters into the analysis. To stay on the conservative side, we used the lowest order parameter $a$ in the charged channel only. The latest and most precise experimental value is \cite{KLOE:2016qvh}

\myeq{0cm}{0cm}{\label{a_KLOE_new} a = -1.095 \pm 0.004.}

\section{Calculation \label{Calculation}}

The details of the calculations with explicit formulas can be found in \cite{Kolesar:2016jwe}. We only summarize the basic steps here, which closely follow the procedure outlined in \cite{Kolesar:2008jr}. 

We start by expressing the charged decay amplitude in terms of 4-point Green functions $G_{ijkl}$, obtained from the generating functional of the QCD currents. We compute at first order in isospin breaking, 
the amplitude then takes the form

\myeq{0cm}{0cm}{\label{Green_f}
	F_\pi^3F_{\eta}A(s,t;u)
		= G_{+-83}-\varepsilon_{\pi}G_{+-33}+\varepsilon_{\eta}G_{+-88} + \Delta^{(6)}_{G_D},}
		
\no where $\Delta^{(6)}_{G_D}$ is the direct higher order remainder to the 4-point Green functions. 
The physical mixing angles to all chiral orders and first order in isospin breaking
can be expressed in terms of quadratic mixing terms of the generating functional to NLO (${M}_{38}^{(4)}$ and $Z_{38}^{(4)}$) and related remainders $\Delta_{M_{38}}^{(6)}$ and $\Delta _{Z_{38}}^{(6)}$

\myeq{0cm}{0cm}{
	\varepsilon_{\pi,\eta} = -\frac{F_{0}^{2}}{F_{\pi^0,\eta}^{2}}
		\frac{({M}_{38}^{(4)}+\Delta_{M_{38}}^{(6)}) - 									
		M_{\eta,\pi^0}^{2}(Z_{38}^{(4)}+\Delta _{Z_{38}}^{(6)})}
		{M_\eta^2-M_{\pi^0}^2}.}
		
In this approximation the neutral decay amplitude can be straightforwardly obtained from the charged one using isospin symmetry and charge conjugation invariance

\myeq{0cm}{0cm}{\label{isospin rel}
	\overline{A}(s,t;u) = -A(s,t;u)-A(t,s;u)-A(u,t;s).}
		
In accord with the method, $O(p^2)$ parameters appear inside loops in the strict chiral expansion, while physical quantities in outer legs. Such a strictly derived amplitude has an incorrect analytical structure due to the leading order masses in loops, cuts and poles being in unphysical positions. We have developed several ways to account for this in \cite{Kolesar:2008jr}, explicit formulas for the $\eta$$\,\to\,$$3\pi$ case are listed in \cite{Kolesar:2016jwe}. The simplest approach is to exchange the LO masses in unitarity corrections and chiral logarithms for physical ones. A more sophisticated method is to use a dispersive representation for the unitarity part of the amplitude 

\myeq{0cm}{0cm}{F_\pi^3 F_\eta A(s,t;u) = \mathcal{P}(s,t;u) + F_\pi^3 F_\eta \mathcal{U}(s,t;u) + O(p^{6}),}

\no where $\mathcal{U}(s,t;u)$ is the unitary part and $\mathcal{P}(s,t;u)$ is an unknown polynomial. This can be obtained by using the reconstruction theorem \cite{Zdrahal:2008bd}, first used in \cite{Knecht:1995tr}. Then the two representations can be sewed together

\myeq{0cm}{0cm}{F_\pi^3 F_\eta A(s,t;u) = G_\mathrm{pol}(s,t;u) + F_\pi^3F_ \eta \mathcal{U}(s,t;u) 
	+ \Delta^{(6)}_{\mathcal{G}_D}}
	
\no where the polynomial part $G_\mathrm{pol}(s,t;u)$ is obtained from (\ref{Green_f}) by the sewing procedure.

However, there is an ambiguity in the derivation of $\mathcal{U}(s,t;u)$. When using the Cutkosky rule, there is a freedom in the way how  to define the relation of the amplitude and the Green function of the entering sub-process at leading order. The most straightforward way is to keep the order by order relation

\myeq{0cm}{0cm}{\label{disp_Fp}
S^{(n)}(s,t;u)=\left( \prod_{i=1}^4F_{P_i}\right)^{-1}G^{(n)}(s,t;u).}

\no Such a definition has the advantage of satisfying perturbative unitarity. On the other hand, a suppression factor $F_0^4/(\prod_i F_{P_i}^2)$ appears in the loop functions, where $P_i$ are the pseudocalars running in the loop. Because we expect the unitarity correction in the case of the $\eta$$\,\to\,$$3\pi$ decays to be sizable, we decided to implement an alternative definition

\myeq{0cm}{0cm}{\label{disp_F0}
S^{(2)}(s,t;u) = F_0^{-4}G^{(2)}(s,t;u).}

\no No suppression factor occurs in this case. Both approaches are valid and they differ in the definition of the higher order remainder $\Delta^{(6)}_{\mathcal{G}_D}$. In our view, we should prefer such a definition where the higher orders are under better control. The approaches will be numerically compared in Section \ref{chi2_analysis}.

The next step is the treatment of the LECs. As discussed above, the leading order
ones, as well as quark masses, are expressed in terms of convenient parameters 

\myeq{0cm}{0cm}{X=X(3),\ \ Z=Z(3),\ \ r,\ \ R.}

At next-to-leading order, the LECs $L_4$-$L_8$ are algebraically reparametrized in terms
of pseudoscalar masses, decay constants and the free parameters $X$, $Z$ and $r$ using chiral expansions of 
two point Green functions, similarly to \cite{DescotesGenon:2003cg}. Because expansions are formally not truncated, each generates an unknown higher order remainder

\myeq{0cm}{0cm}{L_4,L_5,L_6,L_7,L_8\ \to\ \delta_{F_\pi},\delta_{F_K},\delta_{M_\pi},\delta_{M_\eta},\delta_{M_K}.}

We don't have a similar procedure ready for $L_1$-$L_3$ at this point. Therefore we collect
a set of standard $\chi$PT fits \cite{Bijnens:2014lea,Amoros:2001cp,Bijnens:2011tb,Bijnens:1994ie} and by taking their mean and spread, while ignoring the much smaller reported error bars, we obtain an estimate of their influence

\myeq{0cm}{0cm}{L_1^r(M_\rho) = (0.57 \pm 0.18) \cdot 10^{-3}}
\myeq{0cm}{-0.5cm}{L_2^r(M_\rho) = (0.82 \pm 0.28) \cdot 10^{-3}}
\myeq{0cm}{-0.5cm}{L_3^r(M_\rho) = (-2.95 \pm 0.38) \cdot 10^{-3}}

\no These uncertainties enter our statistical analysis. However, as it is discussed in \cite{Kolesar:2016jwe}, the results depend on the value of the constants $L_1$-$L_3$ only very weakly. 

The $O(p^6)$ and higher order LECs, notorious for their abundance, are implicit in
a relatively smaller number of higher order remainders. We have eight indirect remainders
- three generated by the expansions of the pseudoscalar masses, three by the decay constants and two by the mixing angles.
We expand the direct remainder to the 4-point Green functions around the center of the Dalitz plot $s_0=1/3(M_\eta^2$+$2M_{\pi^+}^2$+$M_{\pi^0}^2)$ to second order in Mandelstam variables
		
\myeq{0cm}{0cm}{
	\Delta^{(6)}_{\mathcal{G}_D} = \Delta_A+\Delta_B(s-s_0) + \Delta_C(s-s_0)^2+\Delta_D [(t-s_0)^2+(u-s_0)^2]}
	
\no and thus get four derived direct remainders, two NLO ones ($\Delta_C,\Delta_D$) and two NNLO ones ($\Delta_A,\Delta_B$). We should note that in this approximation we completely miss the analytical structure of the amplitude at higher chiral orders, which includes $\pi\pi$ rescattering effects at NNLO and higher. A deeper discussion of the issue can be found in \cite{Kolesar:2016jwe}. We argue that because the experimental curvature of the Dalitz plot is very small \cite{KLOE:2008ht}, the expansion to second order in the Mandelstam variables is sufficient for the purpose of calculating the decay widths and the lowest order Dalitz slope $a$. On the other hand, the theory seems unreliable for the higher order Dalitz parameters, especially in the case of $b$ and $\alpha$, thus we have chosen to avoid them in this analysis.

For the calculation of the decay widths, we need to numerically integrate over the kinematic phase space. In order to perform the computation efficiently enough, we are forced to expand the amplitude around the center of the Dalitz plot

\myeq{0cm}{0cm}{
	F_\pi^3 F_\eta A(s,t;u) = A+B(s-s_0) + C(s-s_0)^2+D [(t-s_0)^2+(u-s_0)^2].}
	
\no The same argument as above hold here as well - the curvature of the Dalitz plot is tiny, therefore this is a very good approximation for the objective of calculating the decay widths.

\section{$\pi\pi$ scattering \label{sect_pipi_scattering}}

In addition to the $\eta$$\,\to\,$$3\pi$ parameters discussed above, we employ $\pi\pi$ scattering in a very similar way to \cite{DescotesGenon:2003cg}. We use the two lowest order subthreshold parameters in the expansion of the polynomial part of the amplitude, $\alpha_{\pi\pi}$ and $\beta_{\pi\pi}$

\myeqn{-0.5cm}{0cm}{
	A_{\pi\pi}(s,t,u)\ =\ \frac{\alpha_{\pi\pi}}{F_\pi^2}\frac{ M_\pi^2}{3} + 
											\frac{\beta_{\pi\pi}}{F_\pi^2} \left(s-\frac{4 M_\pi^2}{3}\right) +						
											\frac{\lambda_1}{F_\pi^4} \left(s-2 M_\pi^2\right)^2 +
											\frac{\lambda_2}{F_\pi^4} \left[(t-2 M_\pi^2)^2 + (u-2 M_\pi^2)^2 \right] +}
\myeq{2cm}{0cm}{+\frac{\lambda_3}{F_\pi^6} \left(s-2 M_\pi^2\right)^3 +
										\frac{\lambda_4}{F_\pi^6} \left[(t-2 M_\pi^2)^3 + (u-2 M_\pi^2)^3 \right] +
										U_{\pi\pi}^{(4+6)}(s,t;u) + \Op(p^8),}
										
\no where $U_{\pi\pi}^{(4+6)}(s,t;u)$ is the unitary part of the amplitude to NNLO, not given here explicitly. $\alpha_{\pi\pi}$ and $\beta_{\pi\pi}$ can be expressed as
	
\begin{eqnarray}
\alpha _{\pi \pi } &=&1+\frac{3r}{r+2}\epsilon (r)-\frac{2Yr}{r+2}\eta (r)+%
\frac{2(1-X)}{r+2}+\frac{4(1-Y)}{r+2} - \nonumber\\
&&-\frac{1}{2}Y^{2}\left( \frac{M_{\pi }}{4\pi F_{\pi }}\right) ^{2}\left( 
\frac{r}{(r-1)(r+2)}\left( (r+2)\log \left( \frac{M_{\eta }^{2}}{M_{K}^{2}}%
\right) -(r-2)\log \left( \frac{M_{K}^{2}}{M_{\pi }^{2}}\right) \right) +%
\frac{7}{3}\right) -  \nonumber\\
&&-\frac{6}{r+2}\left( \frac{r+1}{r-1}\delta _{M_{\pi }}-\left( \epsilon (r)+%
\frac{2}{r-1}\right) \delta _{M_{K}}\right) 
-Y\frac{2r}{r+2}\left( \frac{r+1}{r-1}\delta _{F_{\pi }}-\left( \eta (r)+%
\frac{2}{r-1}\right) \delta _{F_{K}}\right) + \nonumber\\
&&+2Y\delta _{F_{\pi }}+\delta _{\alpha _{\pi \pi }}
\end{eqnarray}%
\begin{eqnarray}
\beta _{\pi \pi } &=&1+\frac{r\eta (r)}{r+2}+\frac{2(1-Z)}{r+2} + \nonumber\\
&&+\frac{1}{2}Y\left( \frac{M_{\pi }}{4\pi F_{\pi }}\right) ^{2}\left( \frac{%
r}{(r-1)(r+2)}\left( (2r+1)\log \left( \frac{M_{\eta }^{2}}{M_{K}^{2}}%
\right) +(4r+1)\log \left( \frac{M_{K}^{2}}{M_{\pi }^{2}}\right) \right)
-5\right)  - \nonumber\\
&&-\frac{2}{r+2}\left( \frac{r+1}{r-1}\delta _{F_{\pi }}-\left( \eta (r)+%
\frac{2}{r-1}\right) \delta _{F_{K}}\right) +\delta _{\beta _{\pi\pi},}
\end{eqnarray}

\no with

\begin{equation}
Y = \frac{X}{Z},\quad
\epsilon (r)=\frac{2}{r^{2}-1}\left( 2\frac{F_{K}^{2}M_{K}^{2}}{F_{\pi
}^{2}M_{\pi }^{2}}-r-1\right) ,\quad
\eta (r)=\frac{1}{r-1}\left( \frac{F_{K}^{2}}{F_{\pi }^{2}}-1\right). 
\end{equation}	
									
We use the experimental values extracted in \cite{DescotesGenon:2001tn}
										
\myeq{0cm}{0cm}{\alpha_{\pi\pi}^\mathrm{exp} = 1.381 \pm 0.242, \quad \beta_{\pi\pi}^\mathrm{exp} = 1.081 \pm 0.023,\quad \rho_{\pi\pi} = -0.14.}

\no $\rho_{\pi\pi}$ is the correlation coefficient between the two parameters.

\section{Bayesian statistical analysis \label{bayesian_analysis}}

For the statistical analysis, we use an approach based on Bayes' theorem \cite{DescotesGenon:2003cg}

\myeq{0cm}{0cm}{\label{Bayes}
		P(X_i|\mathrm{data}) = \frac{P(\mathrm{data}|X_i)P(X_i)}{\int \mathrm{d}X_i\,P(\mathrm{data}|X_i)P(X_i)}\,,}

\no where $P(X_i|\mathrm{data})$ is the probability density of the parameters and remainders, denoted as $X_i$, having a specific value given the observed experimental data. 

In the case of experimentally independent observables, $P(\mathrm{data}|X_i)$ is the known probability density of obtaining the observed values of the included observables $O_k$ in a set of experiments with uncertainties $\sigma_k$ under the assumption that the true values of $X_i$ are known

\myeq{0cm}{0cm}{
		P(\mathrm{data}|X_i) = \prod_k\frac{1}{\sigma_k\sqrt{2\pi}}\,
				\mathrm{exp}\left[-\frac{(\mathrm{O}_k^\mathrm{exp}-\mathrm{O}_k(X_i))^2}{2\sigma_k^2}\right].}
		
\no Our observables are the charged and neutral decay widths and the Dalitz slope $a$ of the $\eta$$\,\to\,$$3\pi$ decays and the $\pi\pi$ scattering subthreshold parameters $\alpha_{\pi\pi}$ and $\beta_{\pi\pi}$. However, the pair of $\pi\pi$ scattering observables cannot be treated as independent and we need to introduce a correlated probability function for this sector

\myeq{0cm}{0cm}{
	P_{\pi\pi}(\mathrm{data}|X_i) = \frac{1}{2\pi}\sqrt{|C_{\pi\pi}|}\,
	\mathrm{exp}\left(-\frac{1}{2}V_{\pi\pi}^T C_{\pi\pi} V_{\pi\pi}\right),}
	
\no where 	
	
\myeq{0cm}{0cm}{
	V_{\pi\pi} = \left(
		\begin{array}{c}
			\alpha_{\pi\pi}^\mathrm{exp}-\alpha_{\pi\pi}\\ 
			\beta_{\pi\pi}^\mathrm{exp}-\beta_{\pi\pi}
		\end{array}\right), \quad
	C_{\pi\pi} = \frac{1}{1-\rho_{\pi\pi}^2}\left(
		\begin{array}{cc}
			\frac{1}{\sigma_\alpha^2} & \frac{-\rho_{\pi\pi}}{\sigma_\alpha \sigma_\beta} \\
			\frac{-\rho_{\pi\pi}}{\sigma_\alpha \sigma_\beta} & \frac{1}{\sigma_\alpha^2}
		\end{array}\right).}

We resort to Monte Carlo sampling in order to perform the numerical integration in (\ref{Bayes}). It turned out that the uncertainty of latest experimental measurement of the Dalitz parameter $a$ (\ref{a_KLOE_new}) is so small that it is in fact a complication for performing the numerical integration. From this point of view the experimental error is negligible and therefore we can model the experimental distribution as a $\delta$ function rather then a normal distribution

\myeq{0cm}{0cm}{
	\lim_{\sigma_a\to 0}\, \frac{1}{\sigma_a\sqrt{2\pi}}\,
	\mathrm{exp}\left[-\frac{(a_\mathrm{exp}-a)^2}{2\sigma_a^2}\right] =
	\delta(a_\mathrm{exp}-a).} 
	
\no The experimental data thus become a constraint which can be solved algebraically. The solution is most straightforward in the case of the remainder $\delta_B$

\myeq{0cm}{0cm}{\label{delta_a}
	\delta(a_\mathrm{exp}-a) = |K|\,\delta(1+K - \delta_B),}
	
\no where

\myeq{0cm}{0cm}{
		K = \frac{4}{3}\frac{M_\eta(M_\eta-3 M_\pi)}{a_\mathrm{exp}}(1-\delta_A)
				\mathrm{Re}\left(\frac{B}{A}\right).}
				
\no The remainder $\delta_B$ becomes fixed by this constraint and is thus no longer a source of uncertainty.
				
$P(X_i)$ in (\ref{Bayes}) are the prior probability distributions of $X_i$. We use them to implement the theoretical uncertainties connected with our parameters and remainders. In this way we keep the theoretical assumptions explicit and under control. It also allows us to test various assumptions and formulate if-then statements as well as implement additional constraints (see below).

\section{Assumptions \label{assumptions}}

The following list summarizes the higher order remainders we need to deal with: 

\begin{itemize}
		\item[-] $\eta$$\,\to\,$$3\pi$ direct remainders: 
						$\delta_A$, $\delta_B$, $\delta_C$, $\delta_D$ \vsp{-0.25cm}
		\item[-] $\pi\pi$ scattering direct remainders: 
						$\delta_{\alpha_{\pi\pi}}$, $\delta_{\beta_{\pi\pi}}$\vsp{-0.25cm}
		\item[-] indirect remainders: 
						$\delta_{M_\pi}$, $\delta_{F_\pi}$, $\delta_{M_K}$, 
						$\delta_{F_K}$, $\delta_{M_\eta}$, $\delta_{F_\eta}$, 
						$\delta_{M_{38}}$, $\delta_{Z_{38}}$ 
\end{itemize}

\no We use an estimate based on general arguments about the convergence 
of the chiral series \cite{DescotesGenon:2003cg}
		
\myeq{0cm}{0cm}{\label{Delta_G}\delta_G^{(4)}\ \approx\pm 0.3,\quad \delta_G^{(6)}\ \approx\pm 0.1,},

\no where $\delta_G^{(n)}$ collectively denotes all the remainders listed above, with the exception of $\delta_B$, which is fixed by the constraint (\ref{delta_a}) in the main analysis.
We implement (\ref{Delta_G}) by using a normal distribution with $\mu\,$=\,0 and $\sigma\,$=\,0.3 or $\sigma\,$=\,0.1 for the NLO or NNLO remainders, respectively. The remainders are thus limited only statistically, not by any upper bound.

The remainders $\delta_{\alpha_{\pi\pi}}$ and $\delta_{\beta_{\pi\pi}}$, connected with $\pi\pi$ scattering, are defined in such a way that they are of order $O(\hat{m}m_s)$ instead of $O(m_s^2)$. This follows from the particular form of the chiral expansion of the subthreshold parameters $\alpha_{\pi\pi}$ and $\beta_{\pi\pi}$ in the $\hat{m}\to 0$ limit, see \cite{DescotesGenon:2003cg} for details. Therefore we can consider these remainders to be of the order

\myeq{0cm}{0cm}{ 
	\delta_{\alpha_{\pi\pi}}\ \approx\pm 0.03,\quad \delta_{\beta_{\pi\pi}}\ \approx\pm 0.03.}

We assume the strange to light quark ratio $r$ to be known and use the lattice QCD average \cite{Aoki:2013ldr}

\myeq{0cm}{0cm}{r = 27.5 \pm 0.4.}

At last, we are left with three free parameters:

\myeq{0cm}{0cm}{X=X(3),\ Z=Z(3),\ R.}

\no These control the scenario of spontaneous breaking of chiral symmetry and isospin breaking in our results. In the case of $X$ and $Z$, we use a constraint from the so-called paramagnetic inequality \cite{DescotesGenon:1999uh} and assume these parameters to be in the range

\myeq{0cm}{0cm}{\label{prior1}0 < X < X(2),\quad 0< Z < Z(2).}

\no For the two flavor order parameters, we use the lattice QCD values \cite{Bernard:2012fw}. In addition, similarly to \cite{DescotesGenon:2003cg}, we implement constraints following from 

\myeq{0cm}{0cm}{\label{prior2}X(2),Z(2)>\,0, \qquad X/Z=Y<Y_\mathrm{MAX}.}

We use two approaches to deal with $R$. In the first one we assume it to be a known quantity. We use the $N_f$=2+1 lattice QCD average \cite{Aoki:2013ldr} 

\myeq{0cm}{0cm}{R=35.8\pm2.6.}

\no Alternatively, we leave $R$ free, or more precisely, assume it to be in a wide range $R$\,$\in$\,(0,\,80).

\section{Subthreshold parameters of $\pi\pi$ scattering \label{subthreshold_parameters}}

As mentioned in Section \ref{eta3pi_decays}, in \cite{Kolesar:2016jwe} we tested the compatibility of a reasonable convergence of the chiral series, laid out explicitly in the form of assumptions in the previous section, with the experimental data in the case of the $\eta\to 3\pi$ observables. This is an important first step in order to avoid using observables which are problematic from the theoretical or experimental point of view. Analogously, in this section we take a closer look at the subthreshold parameters $\alpha_{\pi\pi}$ and $\beta_{\pi\pi}$, which was not done in \cite{DescotesGenon:2003cg}.

\begin{figure}[h]
	\hsp{-0.25cm}\epsfig{figure=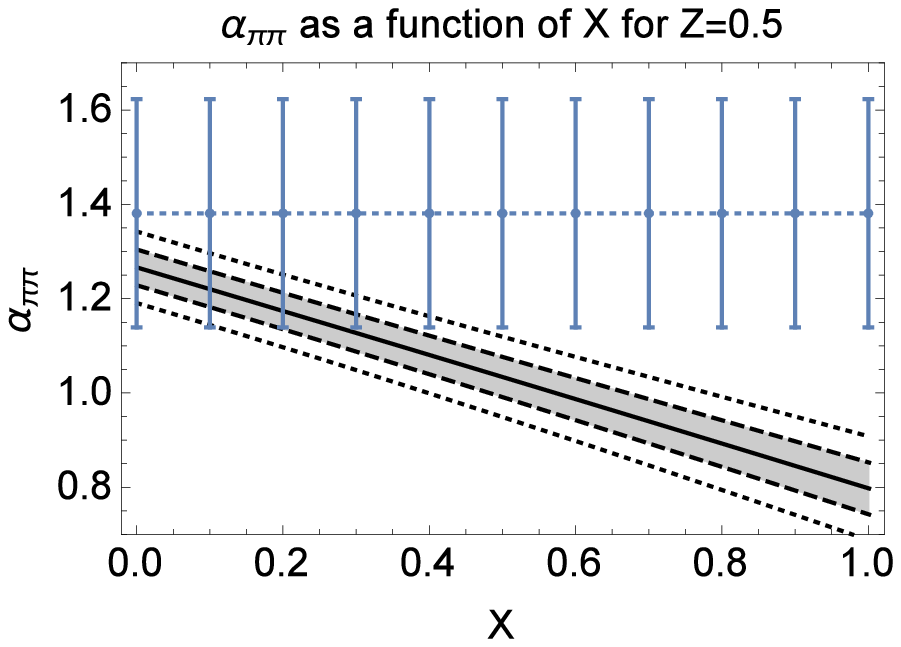,width=0.5\textwidth}
	\hsp{0.5cm}\epsfig{figure=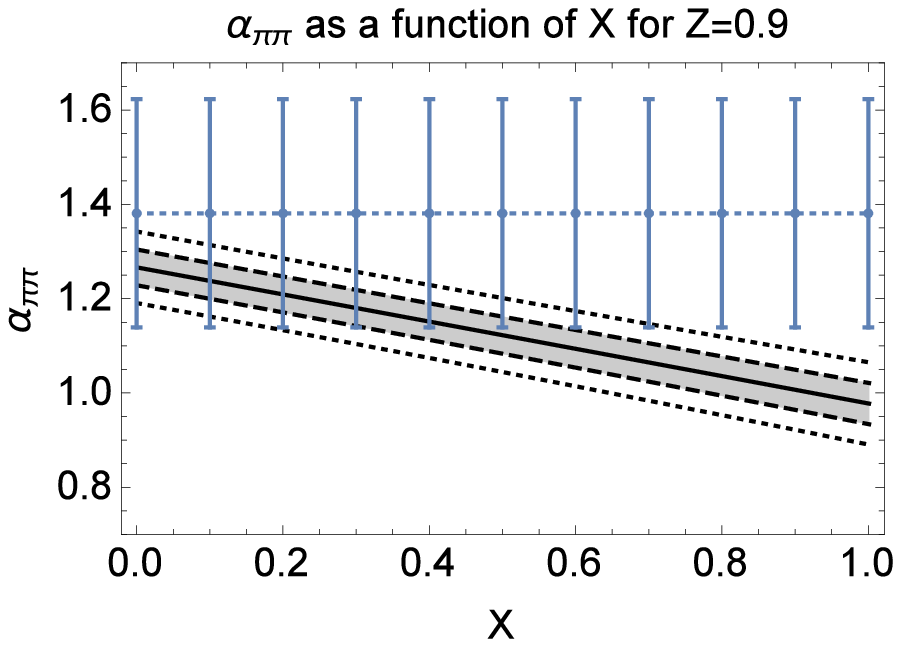,width=0.5\textwidth}\\
	\hsp{-0.25cm}\epsfig{figure=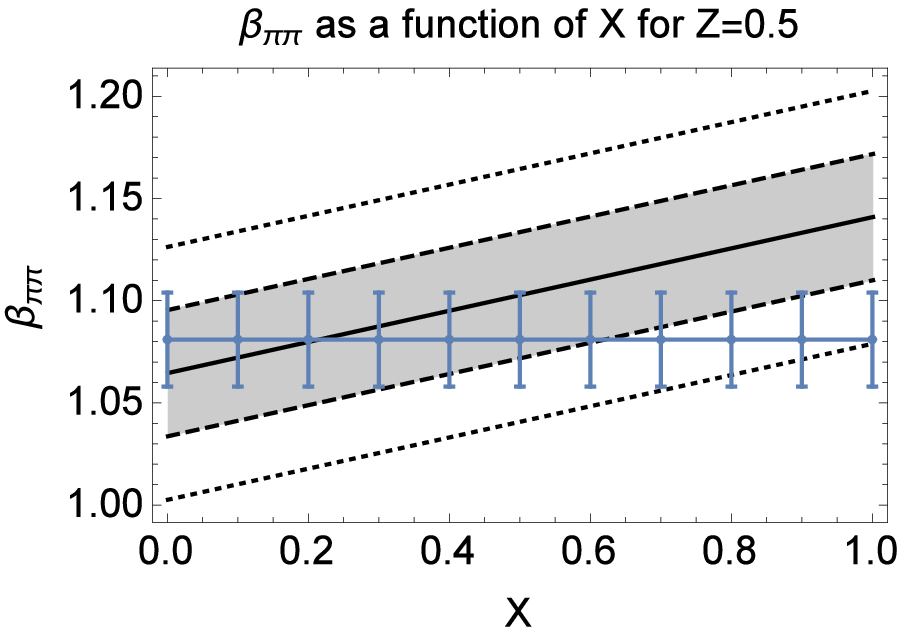,width=0.5\textwidth}
	\hsp{0.5cm}\epsfig{figure=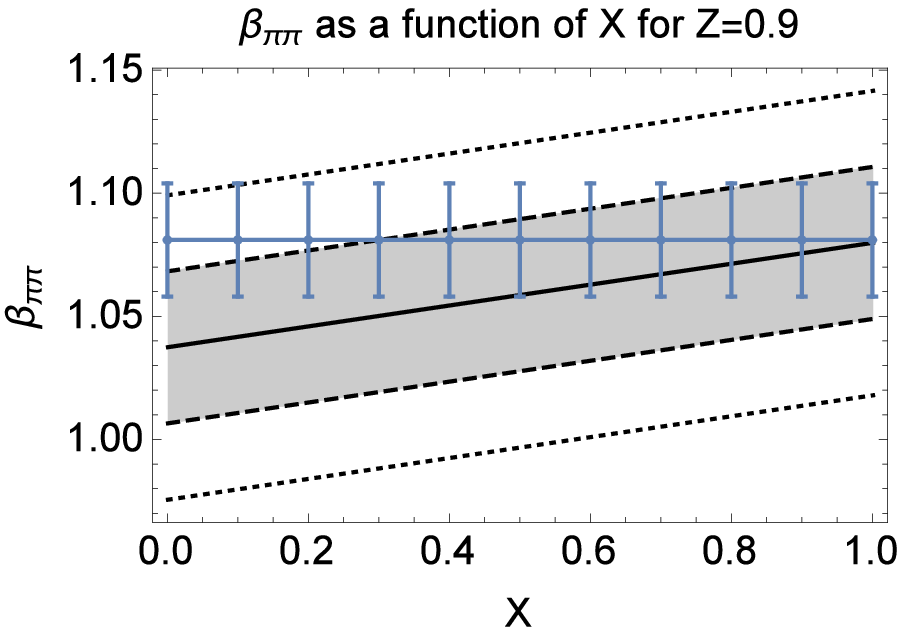,width=0.5\textwidth}
	\caption{Subthreshold parameters $\alpha_{\pi\pi}$ and $\beta_{\pi\pi}$ 
	as a function of $X$ for $Z=0.5 $ and $0.9$. 
	The median (solid line), the one-sigma band (dashed, shadowed) and the two-sigma band (dotted) 	
	are depicted along with the experimental value \cite{DescotesGenon:2001tn}
	(solid horizontal line with error bars) .}
	\label{fig_pipi}
\end{figure}

We numerically generated a large number of theoretical prediction for $\alpha_{\pi\pi}$ and $\beta_{\pi\pi}$, statistically distributed according to the assumptions described in Section \ref{assumptions}. The parameter $Z$ was fixed in two scenarios ($Z$=0.5 and $Z$=0.9), while $X$ was varied in the full range $0<X<1$. Figure \ref{fig_pipi} displays the obtained theoretical distributions in comparison with the experimental data.

As can be seen, both parameters show a significant dependence on $X$, while $\beta_{\pi\pi}$ depends on $Z$ as well. In the case of $\beta_{\pi\pi}$, a broad range of the generated theoretical predictions is consistent with experimental data. Even though the observable seems sensitive to the values of the chiral order parameters, the amount of available information to be extracted might be limited by the experimental error, which is quite substantial.

The picture seems to be more tricky when considering $\alpha_{\pi\pi}$. The central experimental value is outside the 2$\sigma$ band of the theoretical distribution in the whole range of the free parameters. If taken at face value, possibly confirmed by more precise data, it would indicate a very small value of $X$, i.e. a vanishing chiral condensate. Such a scenario would be, however, inconsistent with any current determination of the order parameter (see Tab.\ref{tab2}).
\footnote{A crash of the chiral condensate at three flavors would be also unexpected in the context of an SU(3) gauge theory with a varying number of light quark flavors, see e.g. \cite{Appelquist:2014zsa}.} 
The experimental error on the value of $\alpha$ is very large and our prior expectation therefore is that a large correction of the central value is very possible. It would be thus advisable to be cautious when interpreting the outcomes based on this data in the following.

Our conclusion with regard to the suitability of the subthreshold parameters of $\pi\pi$ scattering for the purpose of extraction of information about the chiral order parameters is hence twofold - both $\alpha_{\pi\pi}$ and $\beta_{\pi\pi}$ seem sensitive to the value of one or both of them, but at present available information is limited due to the quality od experimental data we have at hand.

\section{$\chi^2$ based analysis \label{chi2_analysis}}

In addition to the Bayesian analysis described in Section \ref{bayesian_analysis}, the results of which will follow in Section \ref{Results}, we also perform a search for the minimum of the $\chi^2$ distribution in the Monte Carlo generated set of data points. The aim is to check the quality with which this set of theoretical predictions can reconstruct the experimental data, which is hard to quantify using the Bayesian method. This test also enables us to compare various theoretical approaches, e.g. the alternative ways the dispersive representation can be implemented. 

Because we have much more free parameters than experimental inputs, we generally expect a well working theory to be able to reconstruct the experimental data very precisely with a large enough sample of generated data points. This means the minimum of $\chi^2/n$, where $n$ is the number of experimental observables employed, should be close to zero. On the other hand, a min.$\chi^2/n\approx 1$ means there is typically a 1$\sigma$ deviation between the best of the generated theoretical predictions and the experimental data, which might signal some tension and would not be fully satisfactory. 

As the minimum of the $\chi^2$ distribution is subject to fluctuations stemming from the statistical nature of our procedure, we also report the number of points for which $\chi^2/n<1$. This metric then reveals how well the region of the parameter space where the experimental data lie is covered by the generated theoretical predictions. A reasonably high number should be considered a necessary condition for a well founded analysis, as it demonstrates that the theory has no obvious problem to reconstruct the experimental data and also that we have generated a large enough sample of points.

When comparing different theoretical approaches, one model can overlay the experimental region with fewer points than the other for several reason. It might be that the points originate from a less probable part of the theoretical distribution (e.g., the tail) which means that less likely values of the parameters and remainders are needed in order to reconstruct the experimental data. In that sense, the model is less probable to be true. The $\chi^2$ based analysis can thus form a basis for preference when choosing between alternative approaches. However, it is also possible that one model is simply more sensitive to the values of some of the parameters or remainders, which means the theoretical distribution has a larger spread. This is no reason to exclude to model, of course, and one should be aware of this possibility and check for it.

Let us first explore the issue of deciding between the two dispersive representations used to define the bare expansion of the $\eta\to 3\pi$ amplitude, as discussed in Section \ref{Calculation}. For this purpose we numerically generated a set of $4-6\cdot10^6$ theoretical predictions and constructed a $\chi^2$ distribution. In order to check the consistency of the generated predictions and a complete set of $\eta\to 3\pi$ data, we have not used the constraint (\ref{delta_a}), but rather an experimental value

\myeq{0cm}{0cm}{a = -1.09 \pm 0.02.}

\no This is a slightly older measurement \cite{KLOE:2008ht} with large enough uncertainty for our purpose.

\begin{table}[h] \small \begin{center}
\begin{tabular}{|c|c|c|c|c|}
	\hline \rule[-0.2cm]{0cm}{0.5cm} 
	free parameters & exp.data & disp.aproach & $\sqrt{\mathrm{min.}\chi^2/n}$ & $N(\chi^2/n<1)$ \\
	\hline \rule[-0.2cm]{0cm}{0.5cm} $X$,$Z$ &
	$\Gamma^+$,$\Gamma^0$,$a$ & (\ref{disp_Fp}) & 0.80 & 57 \\
	\hline \rule[-0.2cm]{0cm}{0.5cm} $X$,$Z$ & 
	$\Gamma^+$,$\Gamma^0$,$a$ & (\ref{disp_F0}) & 0.59 & 1048 \\
	\hline \rule[-0.2cm]{0cm}{0.5cm} $Z$,$R$ & 
	$\Gamma^+$,$\Gamma^0$,$a$ & (\ref{disp_Fp}) & 0.65 & 225 \\
	\hline \rule[-0.2cm]{0cm}{0.5cm} $Z$,$R$ & 
	$\Gamma^+$,$\Gamma^0$,$a$ & (\ref{disp_F0}) & 0.46 & 1068\\
	\hline
\end{tabular} \end{center}
	\caption{$\chi^2$ based comparison of the dispersive representations.}
	\label{tab4}
\end{table}\normalsize

As can be seen in Table \ref{tab4}, the approach (\ref{disp_F0}) is able to reconstruct the data more precisely, while the theoretical distributions have a very similar form. This conforms to our intuition from Section \ref{Calculation} and thus, in what follows, we use the representation based on (\ref{disp_F0}) exclusively.

We use the PDG \cite{PDG:2014kda} as the source of input for the values of pseudoscalar masses and decay constants. Because the isospin breaking parameter $1/R$ carries the isospin symmetry breaking, we need to choose a value of these constants in the isospin limit. We use the averaged kaon mass and the well known decay constants of the charged pion and kaon.  However, we found the situation to be quite subtle for the case of the pion mass. For the charged decay $\eta$$\,\to\,$$\pi^+\pi^-\pi^0$ it seems appropriate to use the averaged mass

\myeq{0cm}{0cm}{\overline{M_\pi}^2 = \frac{1}{3}\left(M_{\pi^0}^2+2 M_{\pi^+}^2\right),} 

\no while in the case of the neutral channel $\eta$$\,\to\,$$3\pi^0$ to use the neutral pion mass $M_{\pi^0}$. However, in this approach the isposin relation (\ref{isospin rel}) is not exactly fulfilled. 

Alternatively, one could use the same pion mass for both channels, either the averaged or neutral one, and satisfy the relation (\ref{isospin rel}). One could argue that the difference in the results for the decay widths should be very subtle, of the order $1/R$, and that certainly seems to be true. However, the ratio of the decay widths is known very precisely \cite{PDG:2016xmw}

\myeq{0cm}{0cm}{r_\Gamma\ =\ 1.43 \pm 0.02.}

\no This is an indication that even a slight difference in prediction of order $1/R$ might actually influence whether one can obtain an accurate enough prediction for both the decay widths at the same time. 

As we have found out, this is really the case in the approach with an identical pion mass in both the charged and neutral channel amplitudes, where the predicted ratio $r_\Gamma$ comes out too high. This is reflected in both the minimum of $\chi^2/n$ not being quite close to zero and in the number of points for which $\chi^2/n<1$ to be substantially lower in our $\chi^2$ based test for any value of the parameters in the allowed range, as can be seen in Table \ref{tab5}. Meanwhile, the form of the distributions do not change considerably, as might be expected when only the numerical value of an input parameters is changed. 

\begin{table}[h] \small \begin{center}
\begin{tabular}{|c|c|c|c|c|}
	\hline \rule[-0.2cm]{0cm}{0.5cm} 
	free parameters & exp.data & pion mass & $\sqrt{\mathrm{min.}\chi^2/n}$ & $N(\chi^2/n<1)$ \\
	\hline \rule[-0.2cm]{0cm}{0.5cm} $X$,$Z$ & 
	$\Gamma^+$,$\Gamma^0$,$a$ & $\overline{M_\pi}$, $M_\pi^0$ & 0.59 & 1048 \\
	\hline \rule[-0.2cm]{0cm}{0.5cm} $X$,$Z$ & 
	$\Gamma^+$,$\Gamma^0$,$a$ & $\overline{M_\pi}$ & 0.93 & 7 \\
	\hline \rule[-0.2cm]{0cm}{0.5cm} $X$,$Z$ & 
	$\Gamma^+$,$\Gamma^0$,$a$ & $M_\pi^0$ & 0.92 & 3 \\
	\hline \rule[-0.2cm]{0cm}{0.5cm} $Z$,$R$ & 
	$\Gamma^+$,$\Gamma^0$,$a$ & $\overline{M_\pi}$, $M_\pi^0$ & 0.46 & 1068 \\
	\hline \rule[-0.2cm]{0cm}{0.5cm} $Z$,$R$ & 
	$\Gamma^+$,$\Gamma^0$,$a$ & $\overline{M_\pi}$ & 0.89 & 9 \\
	\hline \rule[-0.2cm]{0cm}{0.5cm} $Z$,$R$ & 
	$\Gamma^+$,$\Gamma^0$,$a$ & $M_\pi^0$ & 0.93 & 3 \\
	\hline
\end{tabular} \end{center}
	\caption{$\chi^2$ based comparison of the pion mass implementations.}
	\label{tab5}
\end{table}\normalsize

In other words, in this case the theory seems to have a harder time to reproduce the experimental data with the required precision. This might look surprising given the number of free parameters in the fit, but one has to realize that while the decay widths in the two channels are independent from the experimental point of view, they are very strongly correlated on the theoretical side given the relation (\ref{isospin rel}). The experiment in fact shows that this relation is not precisely fulfilled in nature. This is the reason we have chosen to implement distinct values of the pion mass in the two channels for the main analysis, which is the same approach as was taken in \cite{Bijnens:2007pr}.

Table \ref{tab6} shows a summary of the $\chi^2$ based tests for our main analysis,  as presented in the next section. This uses the dispersive representation (\ref{disp_F0}), distinct pion mass values for the two $\eta\to 3\pi$ decay modes and the constraint (\ref{delta_a}). The total number of generated data points is $2\cdot10^7$. 

\begin{table}[h] \small \begin{center}
\begin{tabular}{|c|c|c|c|}
	\hline \rule[-0.2cm]{0cm}{0.5cm} free parameters & exp.data & $\sqrt{\mathrm{min.}\chi^2/n}$ & $N(\chi^2/n<1)$ \\
	\hline \rule[-0.2cm]{0cm}{0.5cm} $X$,$Z$ & 
	$\Gamma^+$,$\Gamma^0$,$\delta(a-a_{exp})$ & 0.002 & 165874 \\
	\hline \rule[-0.2cm]{0cm}{0.5cm} $X$,$Z$ & 
	$\Gamma^+$,$\Gamma^0$,$\delta(a-a_{exp})$,$\beta_{\pi\pi}$ & 0.03 & 104670 \\
	\hline \rule[-0.2cm]{0cm}{0.5cm} $X$,$Z$ & 
	$\Gamma^+$,$\Gamma^0$,$\delta(a-a_{exp})$,$\alpha_{\pi\pi}$,$\beta_{\pi\pi}$ & 0.28 & 51278 \\
	\hline \rule[-0.2cm]{0cm}{0.5cm} $Z$,$R$ & 
	$\Gamma^+$,$\Gamma^0$,$\delta(a-a_{exp})$ & 0.003 & 87034\\
	\hline \rule[-0.2cm]{0cm}{0.5cm} $Z$,$R$ & 
	$\Gamma^+$,$\Gamma^0$,$\delta(a-a_{exp})$,$\beta_{\pi\pi}$ & 0.02 & 40919 \\
	\hline \rule[-0.2cm]{0cm}{0.5cm} $Y$ & 
	$\Gamma^+$,$\Gamma^0$,$\delta(a-a_{exp})$,$\alpha_{\pi\pi}$,$\beta_{\pi\pi}$ & 0.15 & 25041 \\
	\hline \rule[-0.2cm]{0cm}{0.5cm} $Y$ & 
	$\Gamma^+$,$\Gamma^0$,$\delta(a-a_{exp})$ & 0.002 & 120130 \\
	\hline \rule[-0.2cm]{0cm}{0.5cm} $Y$ & 
	$\Gamma^+$,$\Gamma^0$,$\delta(a-a_{exp})$,$\beta_{\pi\pi}$ & 0.02 & 62203\\
	\hline \rule[-0.2cm]{0cm}{0.5cm} $Y$ & 
	$\Gamma^+$,$\Gamma^0$,$\delta(a-a_{exp})$,$\alpha_{\pi\pi}$,$\beta_{\pi\pi}$ & 0.28 & 23428\\
	\hline \rule[-0.2cm]{0cm}{0.5cm} $Y$,$R$ & 
	$\Gamma^+$,$\Gamma^0$,$\delta(a-a_{exp})$ & 0.002 & 125724 \\
	\hline \rule[-0.2cm]{0cm}{0.5cm} $Y$,$R$ & 
	$\Gamma^+$,$\Gamma^0$,$\delta(a-a_{exp})$,$\alpha_{\pi\pi}$,$\beta_{\pi\pi}$ & 0.18 & 20054 \\
	\hline
\end{tabular} \end{center}
	\caption{$\chi^2$ based comparison for the main analysis used in Section \ref{Results}.}
	\label{tab6}
\end{table}\normalsize

As can be seen, the number of points seems sufficient. The coverage is better in the approach with $\eta\to 3\pi$ observables only. We can observe a drop in precision when $\alpha_{\pi\pi}$ is included, which correlates with the discussion in Section \ref{subthreshold_parameters}. As we will see, this corresponds to the fact that the signal in the used experimental value of $\alpha_{\pi\pi}$ is in some tension with the one contained in the $\eta\to3\pi$ data.

\section{Results and discussion\label{Results}}

In this section we present the outputs of the Bayesian analysis, i.e. the
probability density functions $P(X_i|\mathrm{data})$,
where $X_{i}\in \{X,Z,Y,R\}$ are the chiral symmetry breaking and explicit
isospin breaking parameters, respectively and $\mathrm{data}$ represent a
subset of the set $\left\{ \Gamma _{+},\Gamma _{0},a,\alpha _{\pi \pi
},\beta _{\pi \pi }\right\} $ of the $\eta \rightarrow 3\pi $ and $\pi \pi
\rightarrow \pi \pi $ observables.

\begin{figure}[t]
	\hsp{-0.25cm}\epsfig{figure=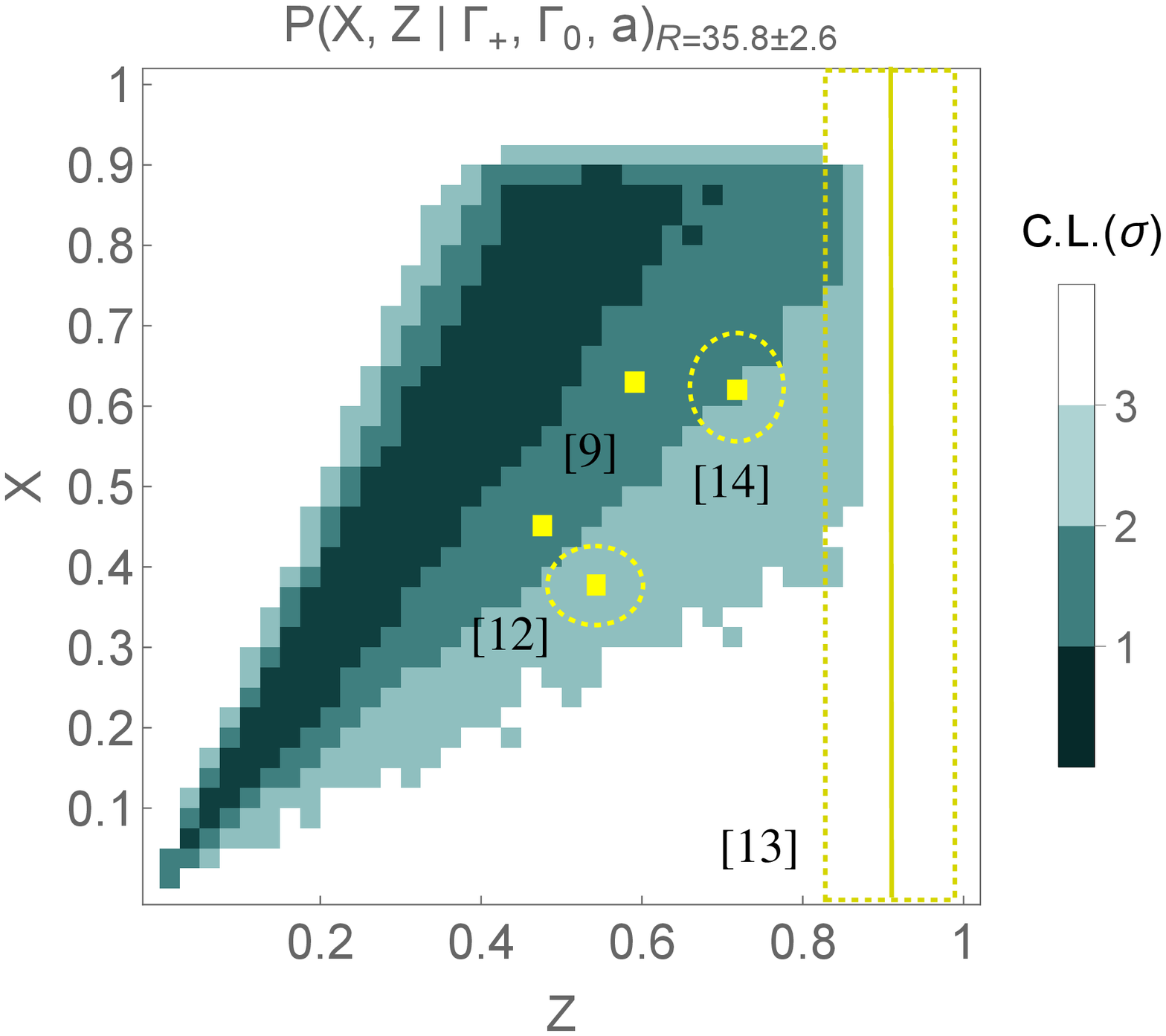,width=0.5\textwidth}
	\hsp{0.5cm}\epsfig{figure=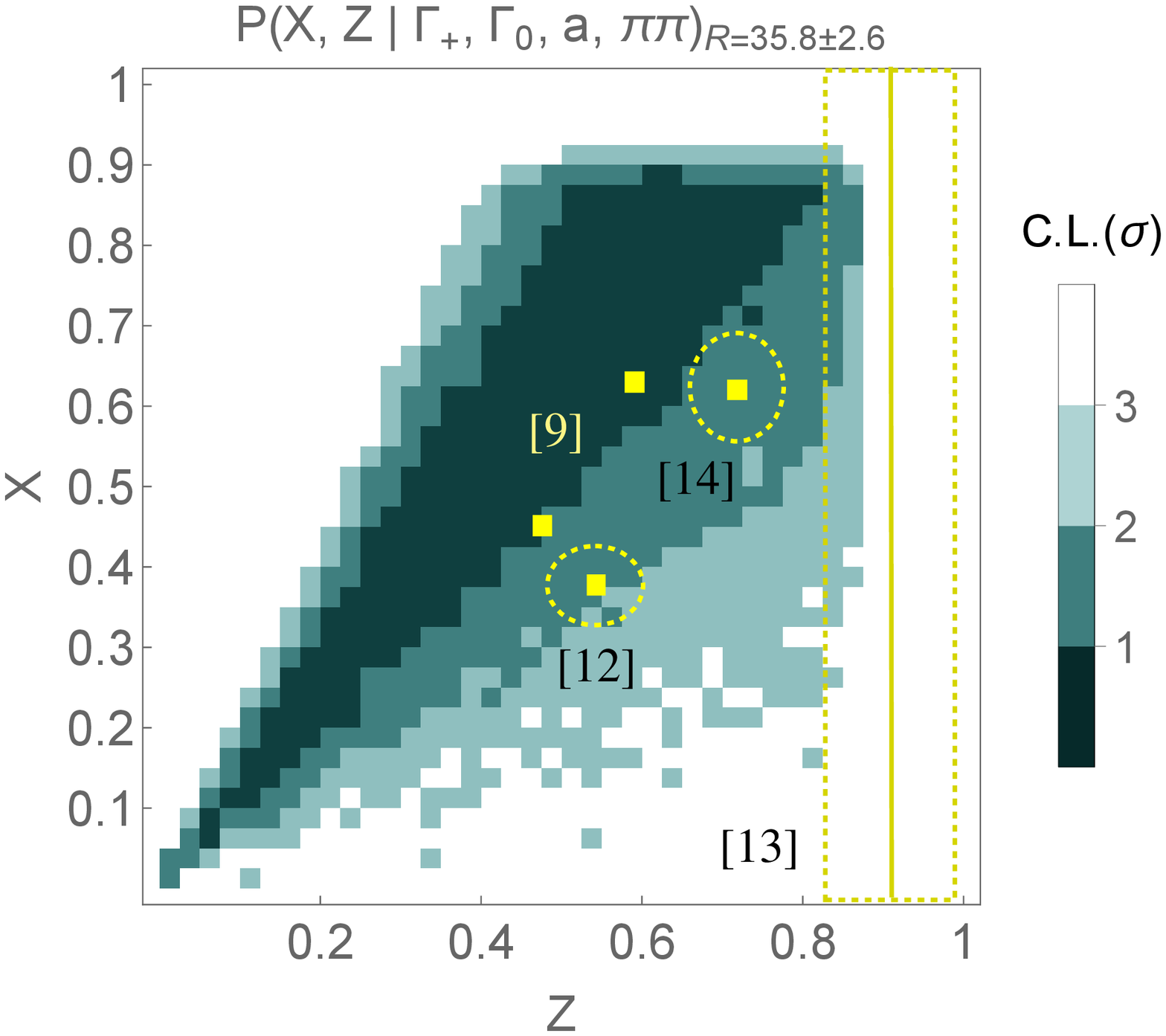,width=0.5\textwidth}
	\caption{Probability density $P(X,Z|\mathrm{data})$ for $R$\,=\,35.8$\,\pm\,$2.6.
					Yellow: results listed in table \ref{tab2}.
					\newline\hsp{1.4cm} Left: $\eta\to3\pi$ data. Right: $\eta\to3\pi$ and $\pi\pi$ scattering data.}
	\label{fig_X-Z}
	\vsp{0.5cm}
	\hsp{-0.25cm}\epsfig{figure=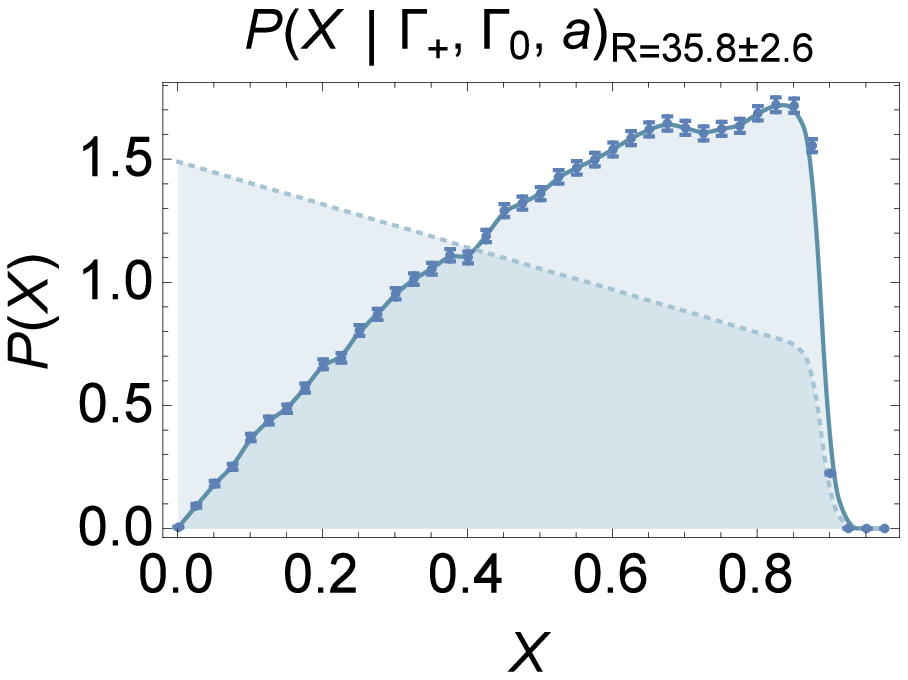,width=0.33\textwidth}
	\hsp{0cm}\epsfig{figure=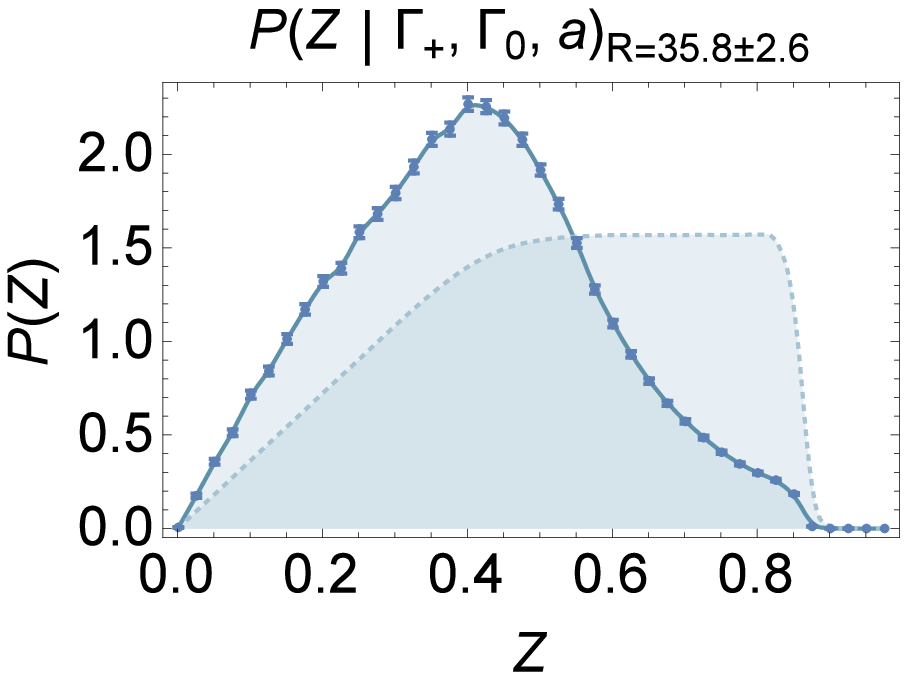,width=0.33\textwidth}
	\hsp{0cm}\epsfig{figure=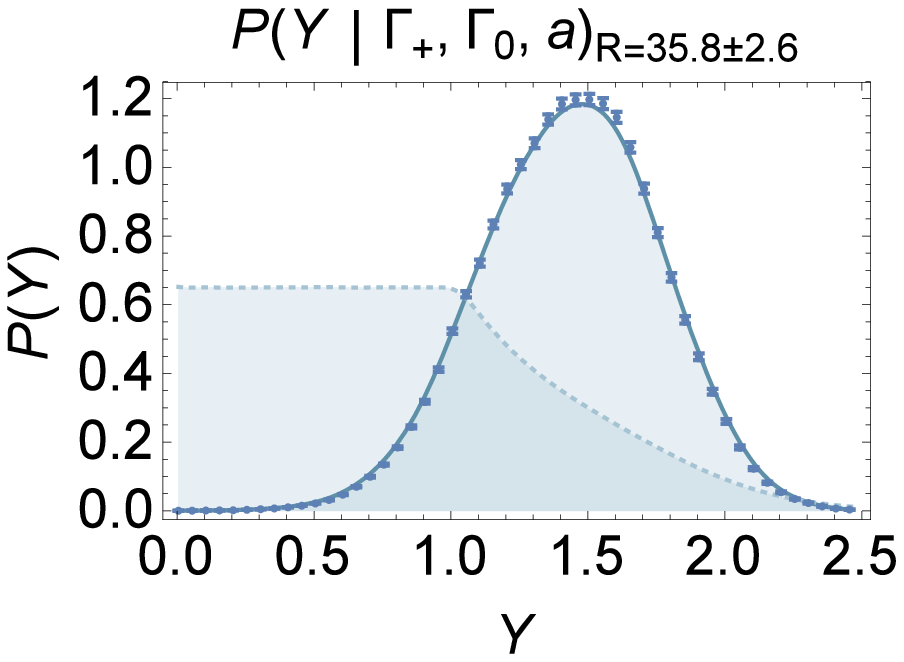,width=0.34\textwidth}
	\caption{Probability densities in comparison with the prior (transparent). 
					$R$\,=\,35.8$\,\pm\,$2.6, $\eta\to3\pi$ data. 
					Smoothed. Error bars estimate the precision of the Monte Carlo integration.}
	\label{fig_1D_R-fixed}
\end{figure}

\begin{figure}[t]
	\hsp{-0.25cm}\epsfig{figure=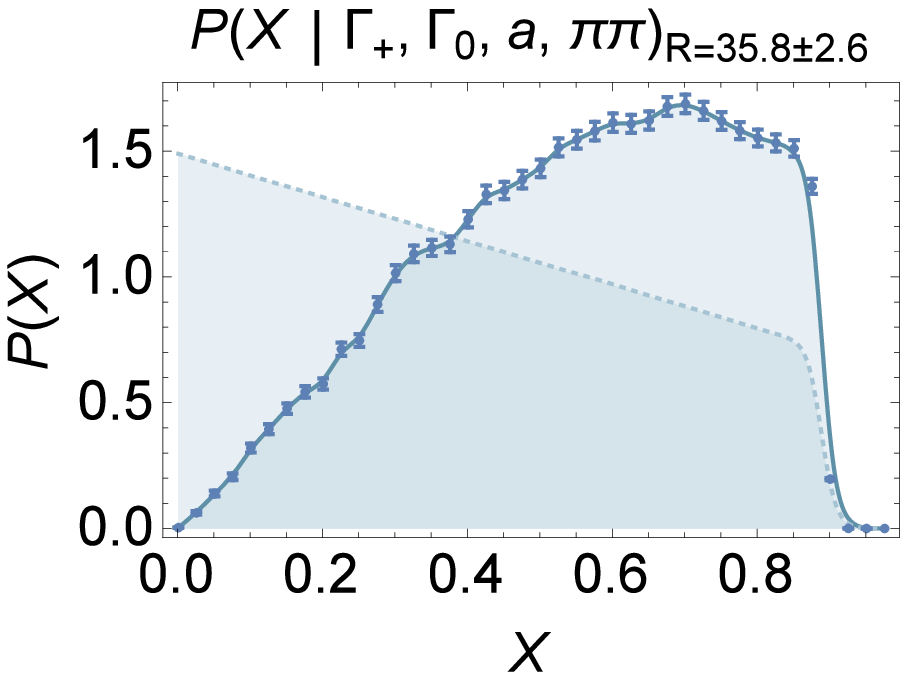,width=0.33\textwidth}
	\hsp{0cm}\epsfig{figure=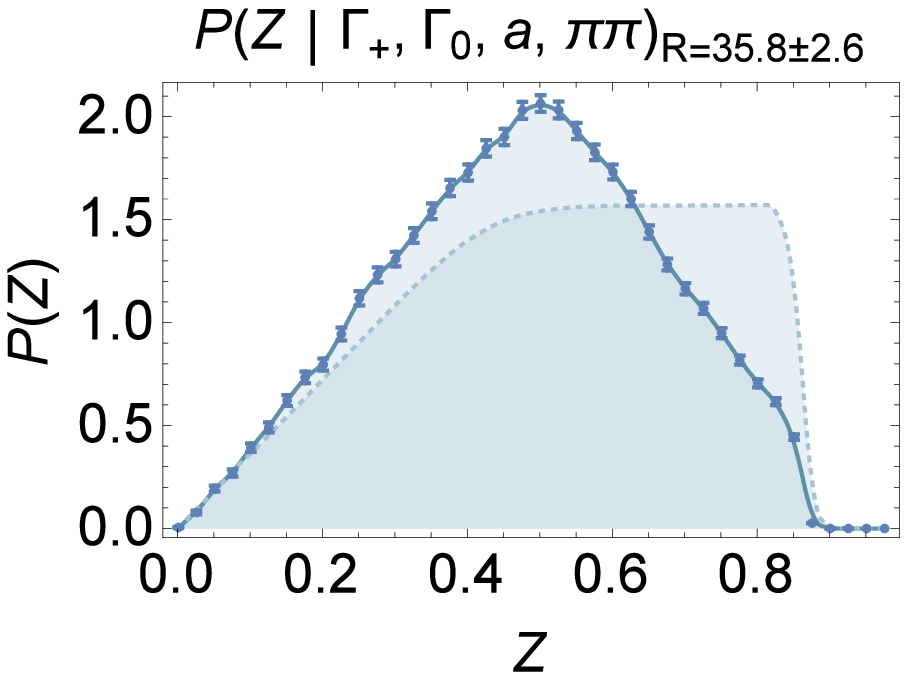,width=0.33\textwidth}
	\hsp{0cm}\epsfig{figure=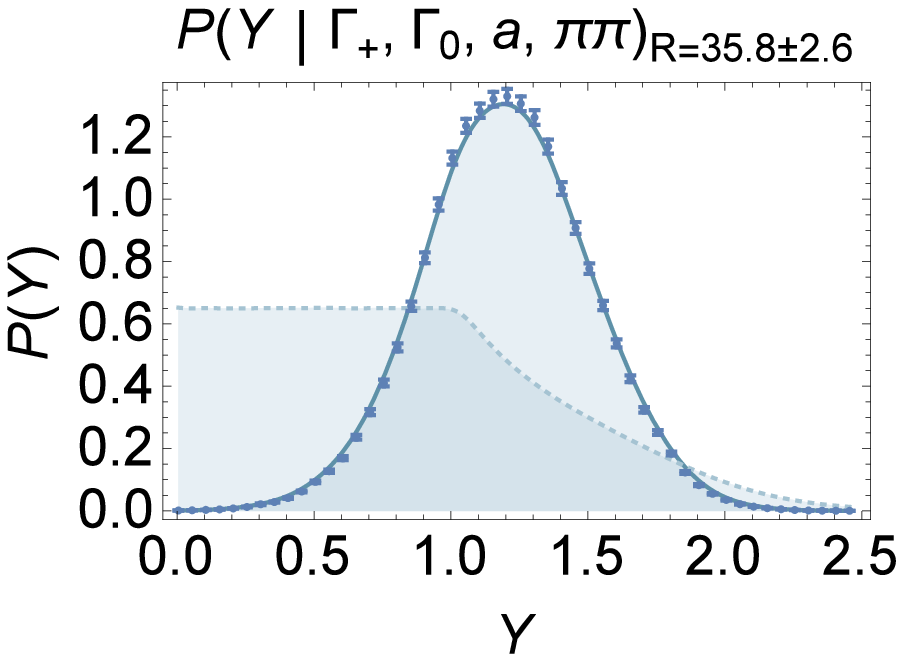,width=0.34\textwidth}
	\caption{Probability densities in comparison with the prior (transparent). 
					$R$\,=\,35.8$\,\pm\,$2.6, $\eta\to3\pi$ and $\pi\pi$ scattering data.
					Smoothed. Error bars estimate the precision of the Monte Carlo integration.}
	\label{fig_1D_R-fixed_pipi}
	\vsp{0.25cm}
	\hsp{-0.25cm}\epsfig{figure=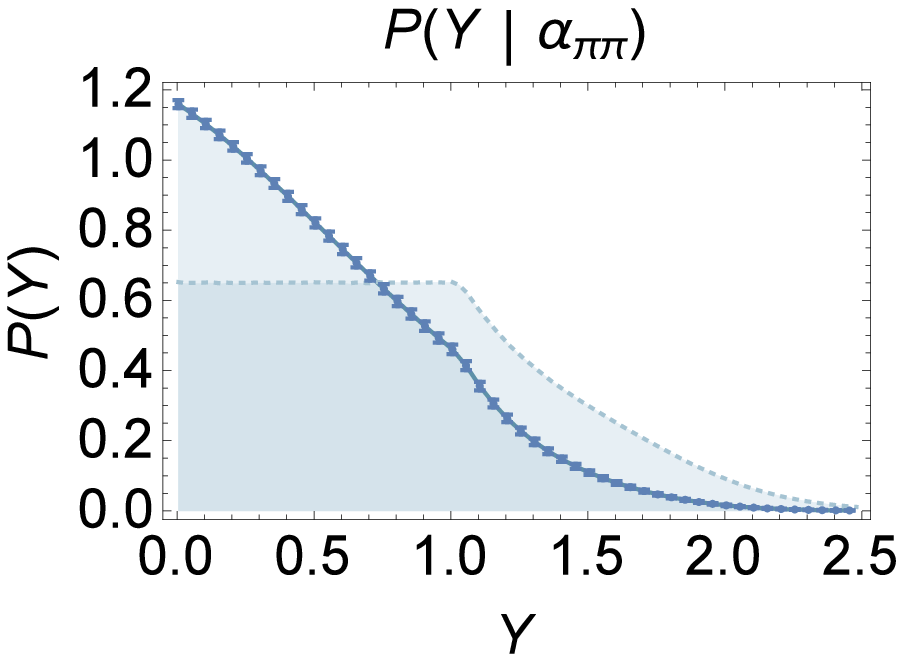,width=0.33\textwidth}
	\hsp{0cm}\epsfig{figure=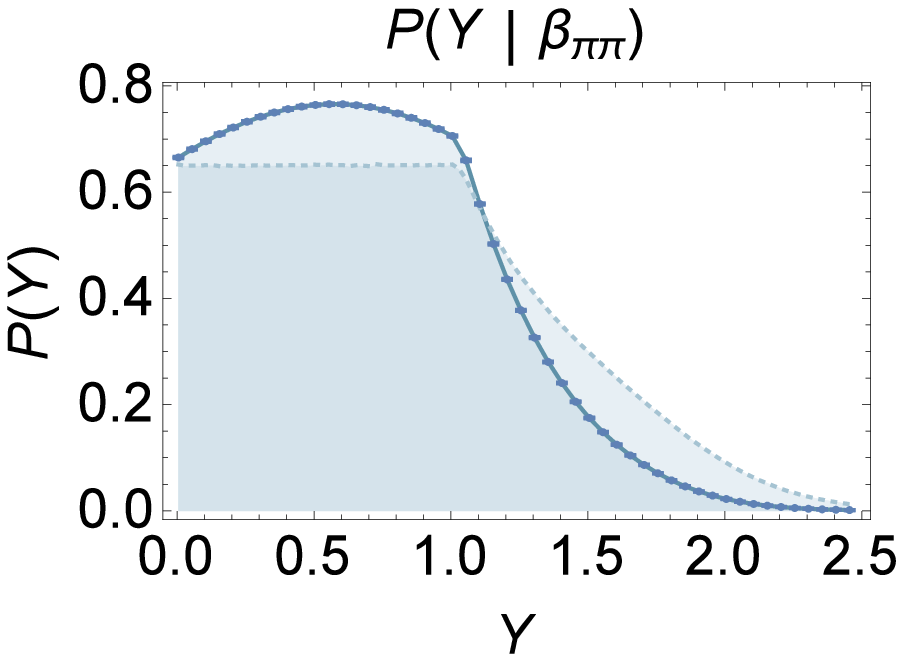,width=0.33\textwidth}
	\hsp{0cm}\epsfig{figure=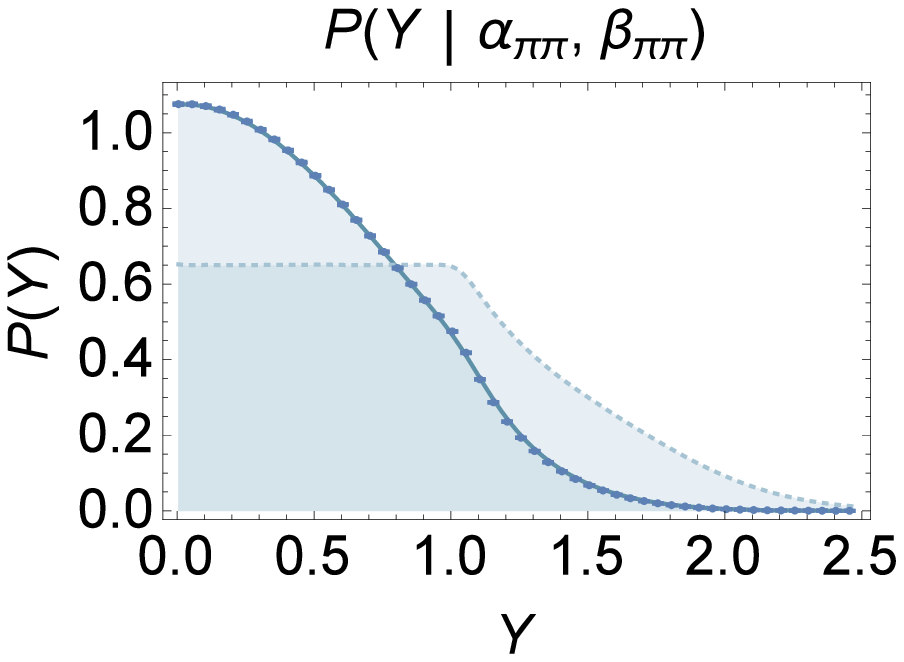,width=0.34\textwidth}
	\caption{Probability densities in comparison with the prior (transparent). 
					$\pi\pi$ scattering data. 
					Smoothed. Error bars estimate the precision of the Monte Carlo integration.}
	\label{fig_1D_pipi}
\end{figure}

\begin{figure}[t]
	\hsp{-0.25cm}\epsfig{figure=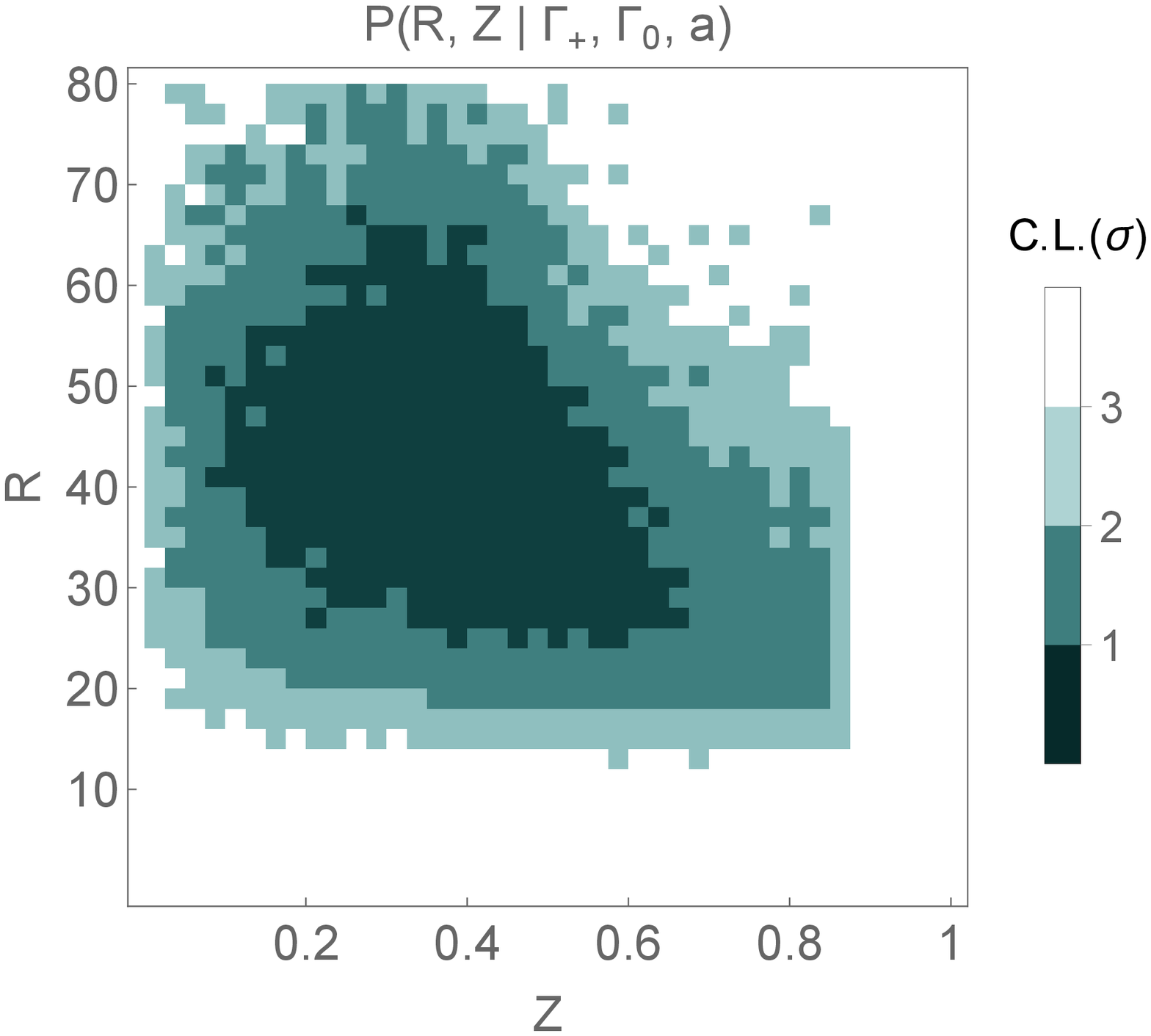,width=0.5\textwidth}
	\hsp{0.5cm}\epsfig{figure=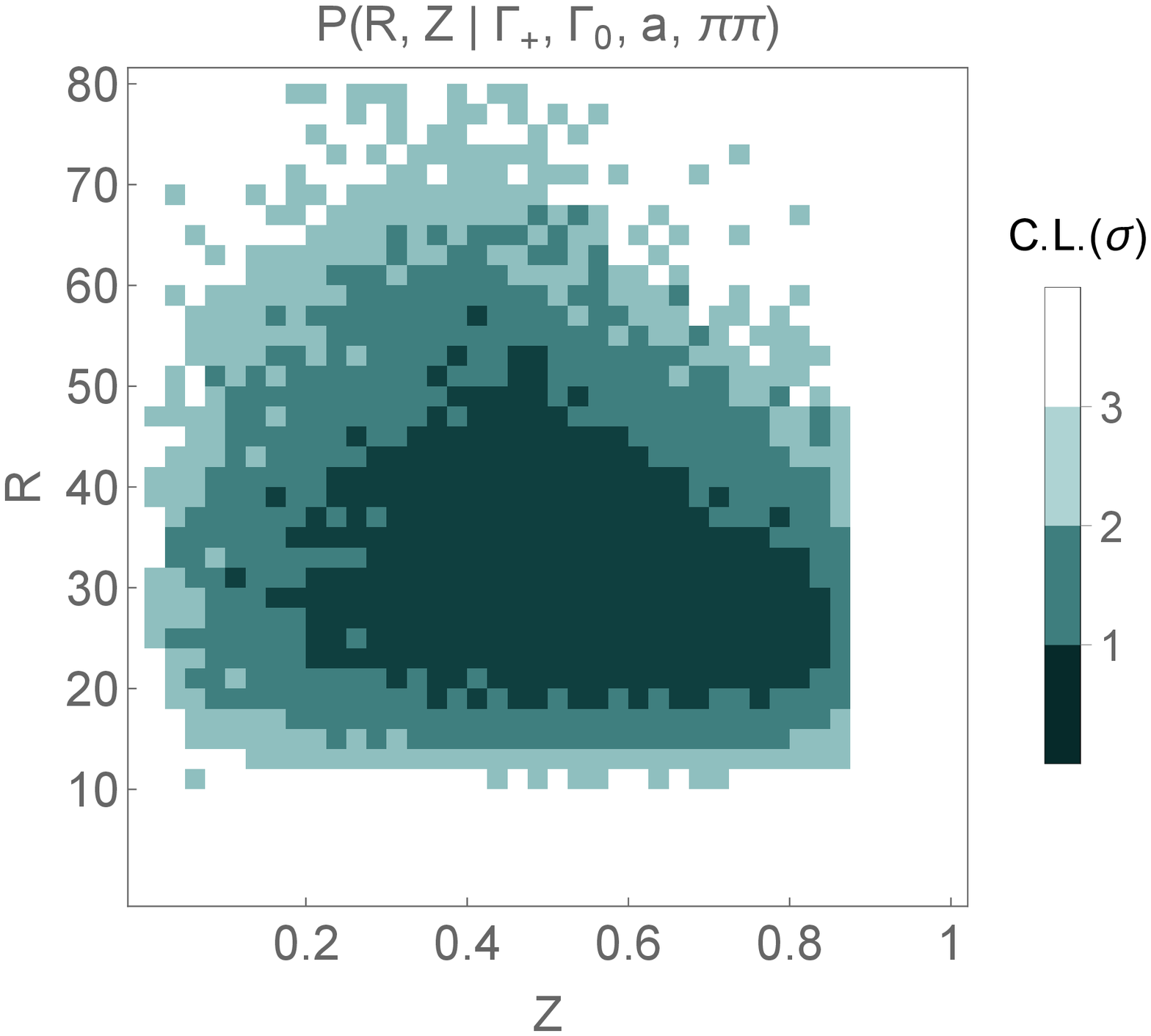,width=0.5\textwidth}
	\caption{Probability density $P(R,Z|\mathrm{data})$ for $R$ free, $X$ integrated out.
					 \newline\hsp{1.4cm} 
					 Left: $\eta\to3\pi$ data. Right: $\eta\to3\pi$ and $\pi\pi$ scattering data.}
	\label{fig_R-Z}
\end{figure}

\begin{figure}[h]
	\hsp{-0.25cm}\epsfig{figure=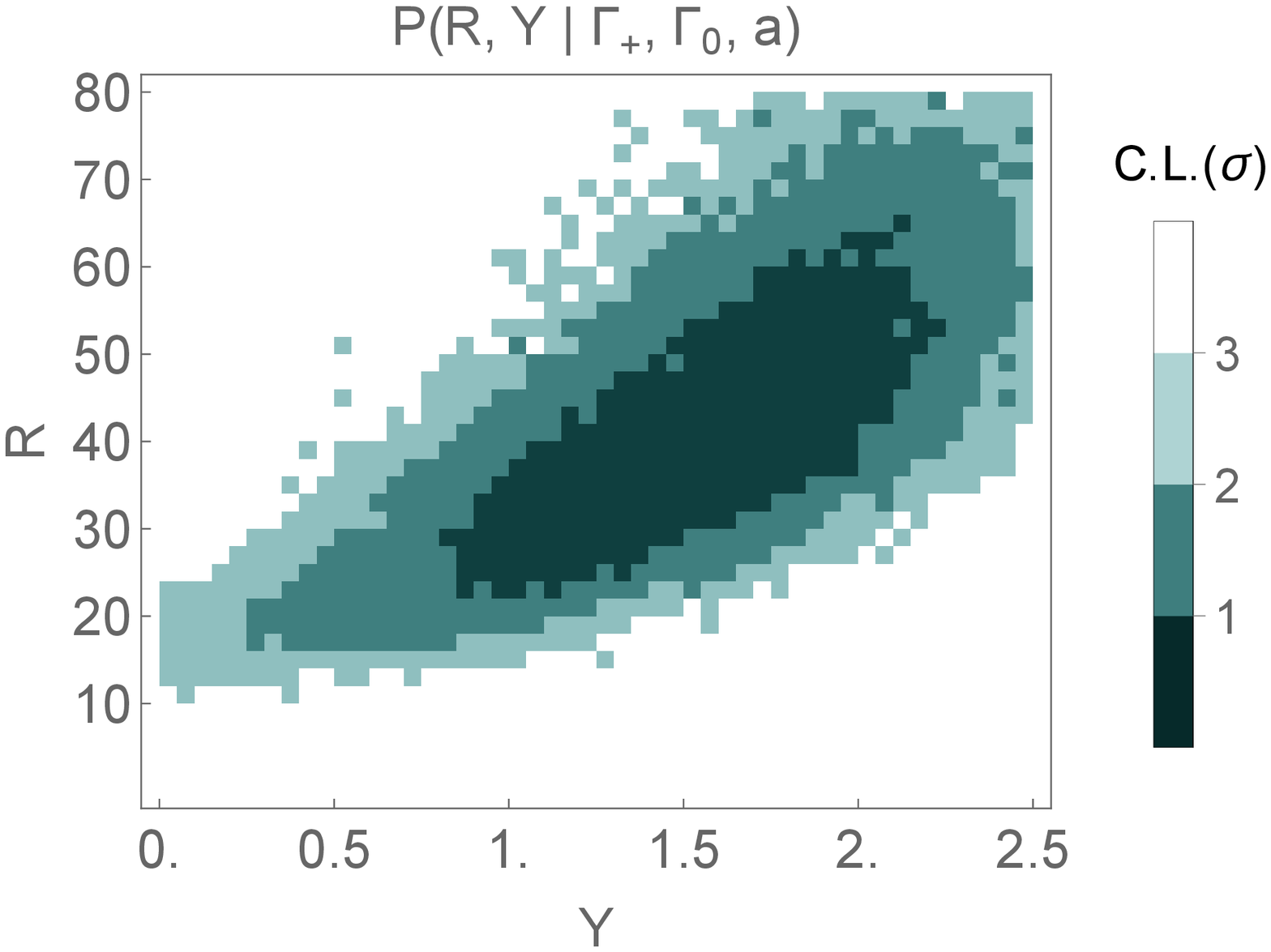,width=0.5\textwidth}
	\hsp{0.5cm}\epsfig{figure=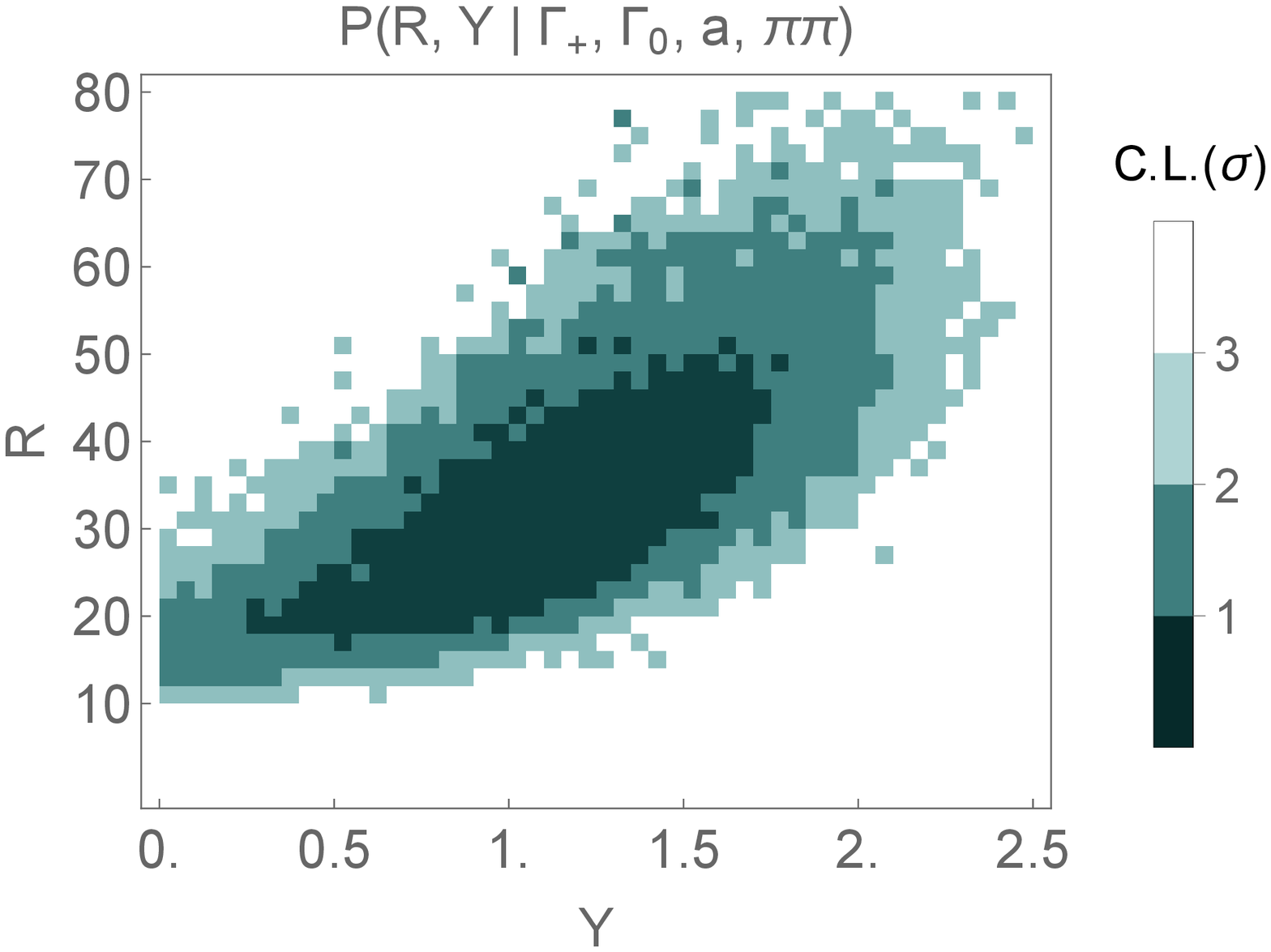,width=0.5\textwidth}
	\caption{Probability density $P(R,Y|\mathrm{data})$ for $R$ free, $Z$ integrated out.
					 \newline\hsp{1.4cm} 
					 Left: $\eta\to3\pi$ data. Right: $\eta\to3\pi$ and $\pi\pi$ scattering data.}
	\label{fig_Y-R}
\end{figure}

\begin{figure}[h]
	\hsp{-0.25cm}\epsfig{figure=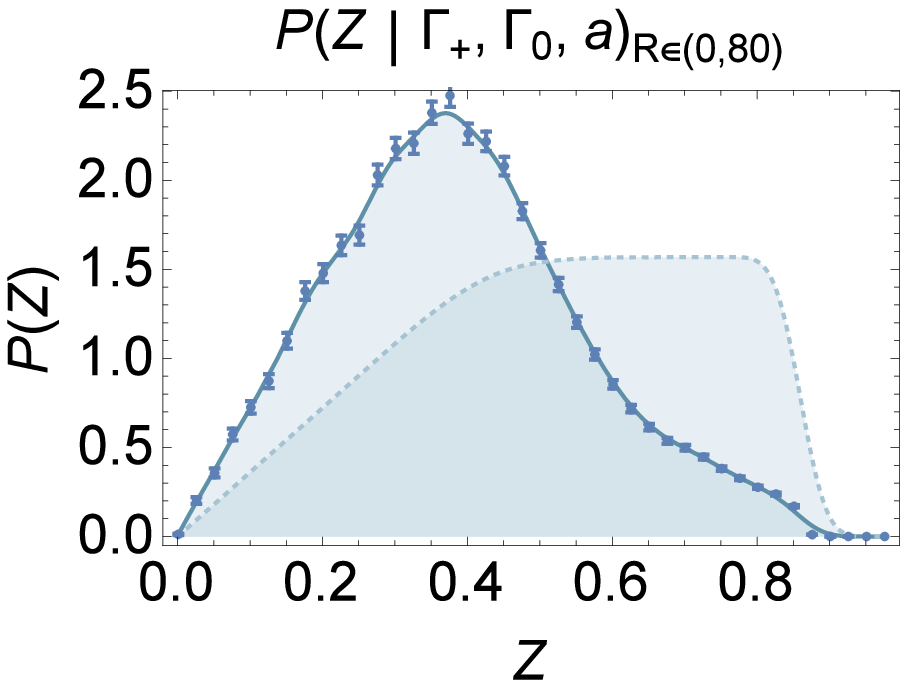,width=0.31\textwidth}
	\hsp{0cm}\epsfig{figure=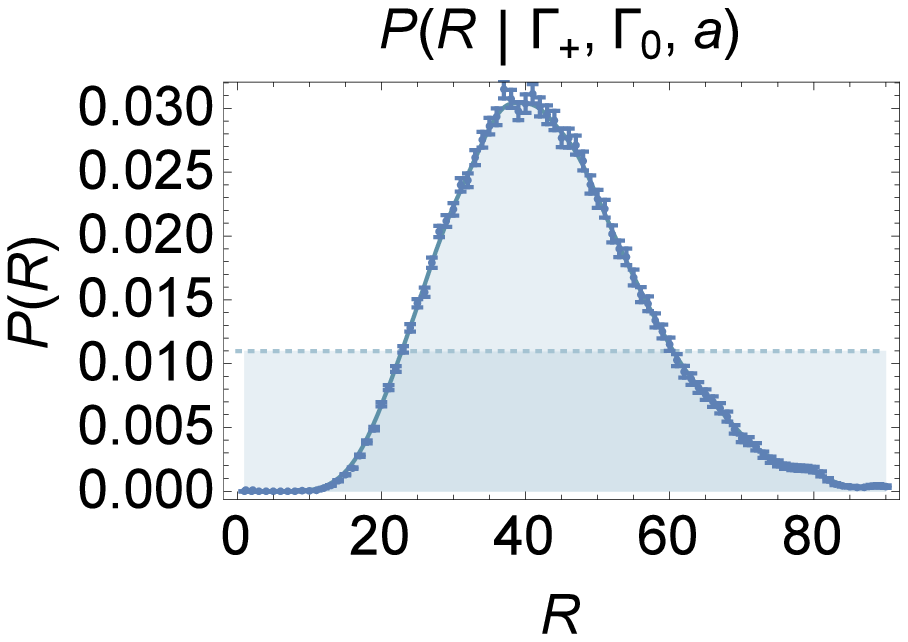,width=0.33\textwidth}
	\hsp{0cm}\epsfig{figure=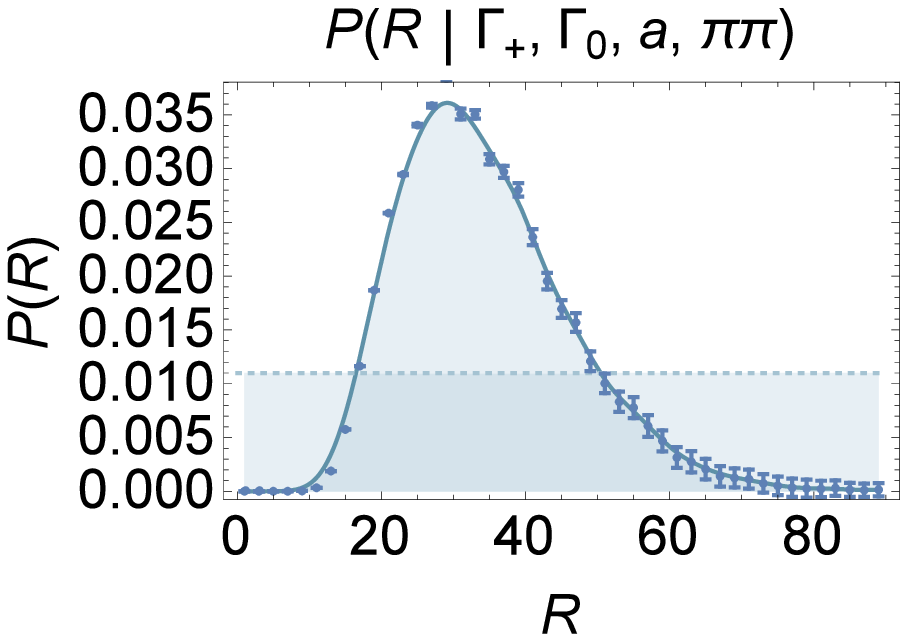,width=0.33\textwidth}
	\caption{Probability densities in comparison with the prior (transparent). 
					$R$ free, $\eta\to3\pi$ data (left, middle), 
					$\eta\to3\pi$ and $\pi\pi$ scattering data (right).
					Smoothed. Error bars estimate the precision of the Monte Carlo integration.}
	\label{fig_1D_R-free}
\end{figure}

The calculated probability density distributions $P(X_i|\mathrm{data})$ are
depicted in Fig.\ref{fig_X-Z}, Fig.\ref{fig_R-Z} and Fig.\ref{fig_Y-R}. 
The colors in these figures highlight the confidence regions of the parameter space with a particular confidence level
\footnote{The choice of such regions could be considered somewhat arbitrary, 
we constructed them by means of integrating the (discretized) probability densities
according to decreasing probabilities starting from the maximal value until the desired 
confidence level was achieved.}. 
The corresponding one dimensional probability density functions for the parameters $X$, $Z$, $Y=X/Z$ and $R$ are obtained by integrating out all other free parameters. In particular, 
\begin{eqnarray*}
P(Y|\mathrm{data}) &=&\int_{0}^{1}\mathrm{d}Z~Z~P(X,Z|\mathrm{data})|_{X\to ZY}.
\end{eqnarray*}%
The results are shown in Fig.\ref{fig_1D_R-fixed}, Fig.\ref{fig_1D_R-fixed_pipi}, Fig.\ref{fig_1D_pipi} and Fig.\ref{fig_1D_R-free}, along with the priors following from the assumptions (\ref{prior1}) and (\ref{prior2}). Error bars in these figures indicate an estimated error of the Monte Carlo integration, which is sufficiently low in all cases. We summarize some characteristics of these distributions in Tab.\ref{tab_R-fixed} and Tab.\ref{tab_R-free}.

\begin{table} \center
\begin{tabular}{|c|c|c|c|c|c|}
\hline
& $\overline{x}$ & $\sigma _{x} $ & median & $%
1\sigma$ ~C.L. & $2\sigma $~C.L. \\ \hline
$P(X|\Gamma _{+},\Gamma _{0},a)$ & 0.56 & 0.22 & 0.58 & (0.31, 0.80) & 
(0.11, 0.88) \\ 
$P(X|\Gamma _{+},\Gamma _{0},a,\pi \pi )$ & 0.56 & 0.21 & 0.58 & (0.32, 0.78)
& (0.13, 0.87) \\ 
$P(Z|\Gamma _{+},\Gamma _{0},a)$ & 0.40 & 0.18 & 0.40 & (0.22, 0.58) & 
(0.08, 0.78) \\ 
$P(Z|\Gamma _{+},\Gamma _{0},a,\pi \pi ) $ & 0.48 & 0.19 & 0.48 & (0.28,
0.68) & (0.11, 0.82) \\ 
$P(Y|\Gamma _{+},\Gamma _{0},a)$ & 1.44 & 0.32 & 1.44 & (1.11, 1.76) & 
(0.78, 2.05) \\ 
$P(Y|\Gamma _{+},\Gamma _{0},a,\pi \pi )$ & 1.20 & 0.30 & 1.20 & (0.90, 1.50)
& (0.60, 1.80) \\ 
$P(Y|\alpha_{\pi\pi} )$ & 0.55 & 0.42 & 0.48 & (0, 0.72) & (0, 1.38) \\ 
$P(Y|\alpha_{\pi\pi},\beta_{\pi\pi} )$ & 0.53 & 0.38 & 0.49 & (0, 0.70) & 
(0, 1.25) \\ \hline
\end{tabular}
\caption{Characteristics of obtained probability distributions, $R$\,=\,35.8$\,\pm\,$2.6.}
\label{tab_R-fixed}\vsp{0.5cm} 
\begin{tabular}{|c|c|c|c|c|c|}
\hline
& $\overline{x}$ & $\sigma _{x} $ & median & $%
1\sigma $ ~C.L. & $2\sigma $~C.L. \\ \hline
$P(R|\Gamma _{+},\Gamma _{0},a)$ & 43.0 & 13 & 41.8 & (29.7, 56.2) & (20.4,
72.2) \\ 
$P(R|\Gamma _{+},\Gamma _{0},a,\pi \pi )$ & 34.4 & 11.6 & 32.7 & (23.0, 45.9)
& (16.7, 62.0) \\
$P(X|\Gamma _{+},\Gamma _{0},a)$ & 0.56 & 0.22 & 0.59 & (0.31, 0.80) & 
(0.10, 0.88) \\ 
$P(Z|\Gamma _{+},\Gamma _{0},a)$ & 0.39 & 0.17 & 0.38 & (0.21, 0.56) & 
(0.08, 0.77) \\ 
$P(Y|\Gamma _{+},\Gamma _{0},a)$ & 1.56 & 0.46 & 1.58 & (1.09, 1.99) & 
(0.57, 2.32) \\ 
 \hline
\end{tabular}
\caption{Characteristics of obtained probability distributions, $R$ free.}
\label{tab_R-free}
\end{table}

\begin{table}[b] \small \begin{center}
\begin{tabular}{|c|c|c|c|}
	\hline \rule[-0.2cm]{0cm}{0.5cm}  & $Z(3)$ & $L_4\cdot10^3$ \\
	\hline 
	\rule[-0.2cm]{0cm}{0.5cm} RBC/UKQCD+large $N_c$ \cite{Ecker:2013pba} & 
			0.91$\pm$0.08 & -0.05$\pm$0.22 \\
	\rule[-0.2cm]{0cm}{0.5cm} NNLO $\chi$PT (BE14) \cite{Bijnens:2014lea} & 
			0.59 & 0.3\\
	\rule[-0.2cm]{0cm}{0.5cm} NNLO $\chi$PT (free fit) \cite{Bijnens:2014lea} & 
			0.48 & 0.76$\pm$0.18 \\
	\rule[-0.2cm]{0cm}{0.5cm} RBC/UKQCD+Re$\chi$PT \cite{Bernard:2012ci} & 
			0.54$\pm$0.06 & 1.06$\pm$0.29\\
	\hline
\end{tabular} \end{center}
	\caption{Correlation of $Z$ and $L_4$ in available determinations of $F_0$.}
	\label{tab_L4}
\end{table}

Let us first discuss the case with the fixed value of $R=35.8\pm 2.6$, which is the lattice QCD average \cite{Aoki:2013ldr} (see Section \ref{assumptions}) and
include only the $\eta \rightarrow 3\pi $ observables into the analysis. This is what we consider our main set of results.  

As can be seen in the left panel of Fig.\ref{fig_X-Z}, a large part of the parameter space can be excluded at 2$\sigma$ CL and the parameters $X$ and $Z$ appear to be quite strongly correlated. The latter statement can be made more quantitative by extracting the result for the ratio of the chiral order parameters $Y = X/Z = 2\hat{m}B_0/M_\pi^2$, see Fig.\ref{fig_1D_R-fixed} and Tab.\ref{tab_R-fixed}, for which we get:

\myeq{0cm}{0cm}{Y = X/Z = 1.44 \pm 0.32.}

\no We also obtain a higher bound for the three flavor chiral decay constant

\myeq{0cm}{0cm}{\label{Z_bound}Z < 0.78\quad \mathrm{(2\sigma\ CL)},}

\no which corresponds to

\myeq{0cm}{0cm}{F_0 < 81\ \mathrm{MeV}\quad \mathrm{(2\sigma\ CL)}.}

\no As can be confirmed from Fig.\ref{fig_1D_R-fixed}, a combination of two factors contribute to this result - the $\eta\to3\pi$ data disfavor high values of $Z$ and the prior (\ref{prior1}), induced by the paramagnetic inequality, causes a sharp cutoff at $Z=Z(2)=0.86\pm0.01$.

As can be seen in Figure \ref{fig_X-Z}, when assuming $R$\,=\,35.8\,$\pm$\,2.6, there is some tension with several of the previous determination of the chiral order parameters (Table \ref{tab2}). The $\eta$$\,\to\,$$3\pi$ data seem to prefer a larger value for the ratio of the order parameters $Y = X/Z$ than recent $\chi$PT and lattice QCD fits. In addition, very large values of the chiral decay constant are excluded at 2$\sigma$ CL. and a relatively small value is favored. The uncertainties, however, are quite large.

Of course, these determinations are hardly compatible among themselves either and provide a very unclear picture. In our view, a reasonable guess could be that there are important differences in how NLO and NNLO low energy constants are treated in various works. In particular, one possible culprit could be the large $N_c$ suppressed LEC $L_4$, which is known to be anti-correlated with $F_0$ for a long time \cite{DescotesGenon:2000ct}. Indeed, this anti-correlation can be observed in the results we quote, see Tab.\ref{tab_L4}. 

In our approach, $L_4$ is reparametrized in terms of the remainders and free parameters,
including $Z$. It can thus vary in a wide range, as we have shown in \cite{Kolesar:2016jwe}.
As the $\eta\to 3\pi$ data seem to constrain $Z$ only mildly, as discussed above, we do not
get significant information on $L_4$ either.  

When concerning \cite{Bijnens:2014lea}, both the main fit (BE14) and the free fit,
which are based on the standard $\chi $PT at $O\left( p^{6}\right) $ and a large set
of the experimental observables, are relatively close to the $2\sigma $ contour of
our distribution. Note however, that the errors of these fits are not at our
disposal. The two fits differ precisely in their treatment of $L_{4}$, 
as indicated in Tab.\ref{tab_L4}. If we roughly estimated the theoretical uncertainty of
these fits as the distance of the corresponding points in the $(X,Z)$ plane, 
then these fits might look quite compatible with our distribution.

The apparent inconsistency with the result of \cite{Bernard:2012ci} is intriguing.
It uses resummed $\chi$PT as well, paired with lattice data. One distinction is a
different approach to the remainders - the authors use uniform distribution of the remainders
inside a closed interval and thus a sharp cutoff. One could speculate that a normal distribution with unbounded tails, as we use, might lead to larger error bars. 

The work \cite{Ecker:2013pba}, which is based on a large $N_c$
motivated approximation of the standard $O\left( p^{6}\right) $ $\chi $PT
calculations, used on lattice data, reports a very large value of the chiral decay constant
and a very low value of $L_4$. This is consistent with the large $N_c$ picture assumed in this paper, but quite far away from other determinations and in tension with our limit (\ref{Z_bound}). In fact, a large part of the region covered by the fit \cite{Ecker:2013pba} is excluded by our prior for $Z$, namely by the constraint stemming from the paramagnetic inequality $Z<Z(2)$ (\ref{prior1}).

Let us now add the $\pi \pi \rightarrow \pi \pi $ data into the analysis. 
As can be seen in the right panel of Fig.\ref{fig_X-Z} and Fig.\ref{fig_1D_R-fixed_pipi}, 
the picture does not change appreciably when including the subthreshold parameters of $\pi\pi$ scattering. Though a bit disappointing, this outcome is not unexpected considering the significant errors connected with the experimental values of these observables and the weak constraints obtained in \cite{DescotesGenon:2003cg} and \cite{DescotesGenon:2007ta}. 

There is one difference, however, we can observe a slight shift of our probability distribution
towards a lower ratio of order parameters $Y=X/Z$, as is confirmed by the mean $\overline{Y}=1.2$ 
($P(Y|\Gamma _{+},\Gamma _{0},a,\pi \pi )$ in Tab.\ref{tab_R-fixed}). This could be interpreted as a move of our predictions in the direction of better compatibility with some of the available determinations (Tab.\ref{tab2}). 
Here we have to be rather cautious though, because, as we have discussed in Sect.\ref{subthreshold_parameters} and Sect.\ref{chi2_analysis}, we can expect some tension between the two sets of data. And indeed, this can be demonstrated when one compares the obtained probability distributions and confidence intervals for $Y$ from $\eta\to 3\pi$  (Fig.\ref{fig_1D_R-fixed}, $P(Y|\Gamma _{+},\Gamma _{0},a)$ in Tab.\ref{tab_R-fixed}) with one from $\pi\pi$ scattering alone (Fig.\ref{fig_1D_pipi}, $P(Y|\alpha_{\pi\pi},\beta_{\pi\pi} )$ in Tab.\ref{tab_R-fixed}), which are barely compatible. We can conclude that we need a more precise determination of the value of the $\pi\pi\to\pi\pi$ subthreshold parameters, especially for 
$\alpha_{\pi\pi}$, to be able to provide a more definite outcome and hopefully solve this puzzle of experimental data pointing in opposite directions.

The results with $R$ left as a free parameter are shown in Fig.\ref{fig_R-Z} and Fig.\ref{fig_Y-R}. The uncertainties are large and thus it's hard to constrain $R$ without additional information on the chiral order parameters and the remainders. Even in this case a part of the parameter space can be excluded at 2$\sigma$ C.L. though.  The obtained value for $R$ (Fig.\ref{fig_1D_R-free}, Tab.\ref{tab_R-free}) is compatible with available results (Table \ref{tab3}). 

We can also evaluate the obtained probability densities for $X$, $Z$ and $Y$ with $R$ left unconstrained (Tab.\ref{tab_R-free}, Fig.\ref{fig_1D_R-free}). Note that dismissing the very clear information on $R$ is not a reasonable assumption, we use it only as a test of robustness. And in this case, the results for $X$ and $Z$ seem almost independent on the value of $R$, which includes the obtained upper bound for the chiral decay constant (\ref{Z_bound}).

On the other hand, as Fig.\ref{fig_Y-R} shows, $R$ and the ratio $Y = X/Z$ seem to be quite strongly correlated. This also provides some additional insight into the large value of $Y$ obtained from $\eta\to 3\pi$ data when fixing $R$\,=\,35.8\,$\pm$\,2.6. Furthermore, as the $\pi\pi$ scattering data we use drag $Y$ down to lower values, one can observe a correlated shift in probability densities of $R$ to smaller numbers in Fig.\ref{fig_Y-R}, Fig.\ref{fig_1D_R-free} and Tab.\ref{tab_R-free}.

\section{Conclusions \label{Conclusions}}

To summarize, we have used statistical methods in the framework of 
resummed chiral perturbation theory to generate large sets of theoretical predictions
for $\eta\to 3\pi$ and $\pi\pi$ scattering observables, dependent on a variety of parameters and assumptions, and confronted them with experimental data.

We have developed a $\chi^2$ based analysis, which allowed
us to form a basis for preference when choosing between alternative assumptions or models. In particular, it showed us that an approach using different pion masses for the charged and neutral
decay channel observables in the isospin limit is significantly more consistent with data,
despite violating the isospin relation, than using identical pion mass for both $\eta\to 3\pi$ decay modes.     

For the main analysis, we have used Bayesian inference to obtain constraints on the values of
three flavor chiral order parameters - the chiral decay constant $F_0$ and the chiral condensate $\Sigma_0$, which are connected with the spontaneous breaking of chiral symmetry in QCD.

When fixing the light quark difference by input from lattice QCD and using $\eta\to 3\pi$ observables only (the decay widths in both channels and the Dalitz plot parameter $a$), we could exclude a large part of the parameters space at 2$\sigma$ CL and have observed some correlation between the chiral order parameters. We have obtained an upper bound for the chiral decay constant, $F_0<81$MeV at 2$\sigma$ CL, and have extracted a fairly large value for the ratio of the order parameters $Y=2\hat{m}B_0/M_{\pi}^2$.

We have found some tension with several of the previous determination of the chiral order parameters, which, however, are neither very consistent with each other. The picture remains unclear, possibly stemming from differences in assumptions about the low energy constants, the large $N_c$ suppressed LEC $L_4$ being one candidate.

The picture have not changed appreciably when we performed a combined $\eta\to 3\pi$ and $\pi\pi\to\pi\pi$ analysis. However, we have observed some tension between $\eta\to 3\pi$ and $\pi\pi$ scattering data, which limited our ability to draw more definite conclusions. The possible source is the experimental error of the observables we used, the subthreshold parameters $\alpha_{\pi\pi}$ and $\beta_{\pi\pi}$. Specifically, the value of $\alpha_{\pi\pi}$ proved suspicious in our investigation, as it prefers a very low value for the chiral condensate, which is not very consistent with current expectations.  

We have also tried to extract information on the difference of light quark masses, but the uncertainties proved to be very large. The result is consistent with available data though.

\emph{Acknowledgment:} This work was supported by the Czech Science Foundation (grant no. GACR 15-18080S).

\bibliographystyle{utphys}
\bibliography{Bibliography}

\end{document}